\documentclass[twocolumn]{jpsj2}

\title{%
DC Josephson Effect in a Tomonaga-Luttinger Liquid}

\author{%
Yositake \textsc{Takane}}

\inst{%
Department of Quantum Matter, Graduate School of Advanced Sciences of Matter,
Hiroshima University, Higashi-Hiroshima 739-8530, Japan
}

\abst{%
The dc Josephson effect in a one-dimensional Tomonaga-Luttinger (TL)
liquid is studied on the basis of two bosonized models.
We first consider a TL liquid sandwiched between
two superconductors with a strong barrier at each interface.
Both the interfaces are assumed to be perfect
if the barrier potential is absent.
We next consider a TL liquid with open boundaries,
weakly coupled with two superconductors.
Without putting strong barriers,
we instead assume that the coupling at each interface
is described by a tunnel junction.
We calculate the Josephson current in each model,
and find that the two models yield same results.
The Josephson current is suppressed by
repulsive electron-electron interactions.
It is shown that the suppression is characterized by
only the correlation exponent for the charge degrees of freedom.
This result is inconsistent with a previously reported result,
where the spin degrees of freedom also affects the suppression.
The reason of this inconsistency is discussed.}

\begin{document}

\maketitle

\section{Introduction}

Progress in micro-fabrication technology has enabled us to
prepare normal-conductor--superconductor (NS) composite systems
without a Schottky barrier at the NS interface.~\cite{rf:1}
A highly controllable normal segment has been realized
by using a two-dimensional electron system
formed at semiconductor heterostructures.~\cite{rf:2}
We are interested in how the electronic transport properties
are influenced by superconducting proximity effect.
Particular attention has been focused on the low-temperature transport
properties with the phase coherence of electrons in the normal segment.
Various experiments have been performed to reveal
unusual features of the conductance in NS systems
and the Josephson effect in SNS systems.~\cite{rf:2, rf:3}

Progress in micro-fabrication technology has also made it possible
to prepare a very narrow quantum wire which can be regarded as
an idealistic one-dimensional (1D) electron system.~\cite{rf:4, rf:5}
The electron transport in a 1D system with a few barriers
has been studied extensively.~\cite{rf:6, rf:7, rf:8, rf:9,
rf:10, rf:11, rf:12, rf:13, rf:14}
The central problem addressed there is how mutual electron-electron
interactions affect the transport properties.
Electron-electron interactions greatly affect the low-energy properties
of a 1D electron system.
A 1D interacting electron system with a gapless excitation is generally
described as a Tomonaga-Luttinger (TL) liquid.~\cite{rf:15, rf:16, rf:17}
The strength of electron-electron interactions is
characterized by the correlation exponents $K_{\rho}$ and $K_{\sigma}$,
where $K_{\rho}$ ($K_{\sigma}$) is related to
the charge (spin) degrees of freedom.
For example, $K_{\rho} < 1$ and $K_{\sigma} = 1$ corresponds to
the spin-independent repulsive interaction cases,
and the system is reduced to the noninteracting electron gas when
$K_{\rho} = K_{\sigma} = 1$.
Reflecting the nature of the TL liquid,
we expect that the electron transport in a 1D system shows an anomalous
property, which is not seen in a conventional Fermi liquid.
Kane and Fisher studied the electron transport in a spin-less TL liquid
with a single barrier potential of $\delta$-function type.~\cite{rf:6}
Using renormalization-group argument, they found that
for repulsive electron-electron interactions with $K_{\rho} < 1$,
the potential becomes large after renormalization even
if it initially is very small.
This means that electrons at low energy see the potential
as if it were effectively infinite.
Thus, the low-energy properties are described by the open-boundary fixed point,
at which the 1D system is effectively disconnected at the barrier.
This anomalous behavior strongly affects the transport through the barrier.
It is shown that the conductance $G$ is suppressed with decrease of
temperature $T$ as $G \propto T^{2K_{\rho}^{-1} - 2}$,
and eventually vanishes at $T = 0$.
This ia a clear manifestation of the TL-liquid behavior.
The extention to spin-dependent cases is achieved
by Kane and Fisher,~\cite{rf:8} and Furusaki and Nagaosa.~\cite{rf:10}
The conductance in this case also obeys a power-law behavior
as a function of $T$,
and its exponent is determined by $K_{\rho}$ and $K_{\sigma}$.
Furthermore,
the electron transport in a TL liquid with a double-barrier structure
was also studied by these authors.~\cite{rf:7, rf:8, rf:9}

Inspired by the developments mentioned above,
considerable attention has been attracted to the transport properties of
a TL liquid coupled with superconductors.~\cite{rf:18, rf:19,
rf:20, rf:21, rf:22, rf:23, rf:24, rf:25, rf:26, rf:27}
The central problem is how the superconducting proximity effect
in a 1D system is modified by electron-electron interactions.
Particular interest is focused on the Josephson effect in
a superconductor--TL-liquid--superconductor (STLLS) system,~\cite{rf:19,
rf:20, rf:21, rf:24, rf:26, rf:27}
which is the issue of the present paper.

\begin{figure}[btp]
\begin{center}
\includegraphics[height=6cm]{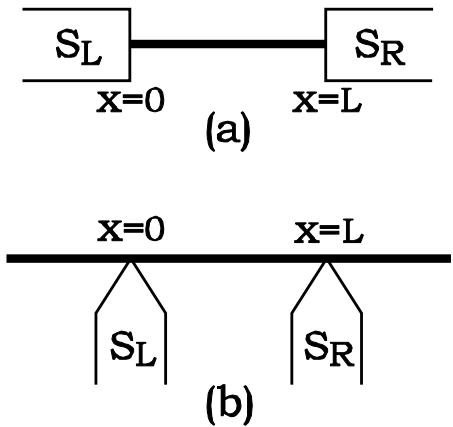}
\end{center}
\caption{
The geometries discussed in the text.
(a) One-dimensional electron system of length $L$ sandwiched
between two superconductors.
(b) One-dimensional electron system of infinite length
weakly coupled with two superconductors.
$\rm S_{L}$ ($\rm S_{R}$) represnts the left (right) superconductor.
}
\label{fig:1}
\end{figure}
Let us consider a clean 1D noninteracting electron system of length $L$
sandwiched between superconductors (see Fig. 1(a)).
We focus our attention on the long-junction case,
where $v_{\rm F}/L$ ($v_{\rm F}$: Fermi velocity) is much smaller than
the energy gap $\Delta$ of superconductors.
In the low-temperature regime of $T \ll v_{\rm F}/L$, the Josephson critical
current obeys $j_{\rm c} \propto e v_{\rm F}/L$.~\cite{rf:28}
In the case of an STLLS system,
we expect that the critical current deviates from
the inversely linear $L$-dependence due to the TL-liquid behavior.
This problem is first discussed by Fazio \textit{et al.}~\cite{rf:19, rf:20}
They considered a clean TL liquid of infinite length weakly connected to
two superconductors (see Fig. 1(b)),
where the separation between the two NS contacts is $L$.
They found that when $T \ll v_{\rm F}/L$,
the $L$-dependence of the critical current is given by
$j_{\rm c}^{\rm FHO} \propto (1/L)^{K_{\rho}^{-1} + K_{\sigma} - 1}$.
This power-law $L$-dependence is a clear manifestation of
the TL-liquid behavior.
It indicates that as long as $K_{\rho}^{-1} + K_{\sigma} > 2$,
the critical current is suppressed by electron-electron interactions
compared with the noninteracting case.
This result shows that both the charge and spin degrees of freedoms
influence the suppression of the critical current.
It should be noted that in calculating the critical current
in the STLLS system,
Fazio \textit{et al.} completely neglected potential scattering
at the contacts between the TL liquid and the superconductors.
Fabrizio and Gogolin~\cite{rf:29} pointed out that
since the potential induced by the contacts becomes large
after renormalization even if it initially is very small,~\cite{rf:6}
the TL liquid is finally disconnected at the NS contacts.
They suggested that in such a situation,
we should employ a TL liquid with open boundaries
to correctly take account of the potential scattering
in studying the Josephson effect.
Maslov \textit{et al.}~\cite{rf:21} studied a different model,
in which a TL liquid of length $L$ is sandwiched between superconductors
(see Fig. 1(a)).
They presented a usefull bosonized description of the STLLS system,
and calculated the Josephson current when the NS interfaces are perfect.
They also examined the $L$-dependence of the critical current
in the presence of a strong barrier at each NS interface.
However, they do not correctly take account of NS boundary condition
in their renormalization-group argument.
The works mentioned above treat a TL liquid with
repulsive electron-electron interactions.
The Josephson effect in attractive interaction cases has been
studied extensively by Affleck \textit{et al.}~\cite{rf:26}

In this paper, we study the dc Josephson effect in a TL liquid
using two different models.
We first consider a TL liquid of length $L$ sandwiched between
two superconductors with a barrier at each NS interface.
Both the NS interfaces are assumed to be perfect
if the barrier potential is absent.
We call it the interface-barrier model.
Particular attention is focused on the limit
where the barrier potential is very strong.
This model is equivalent to that treated by
Maslov \textit{et al.}~\cite{rf:21}
We next consider a TL liquid of length $L$ with open boundaries,
which is weakly connected at its left (right) end
with the left (right) superconductor through a tunnel junction.
We call it the weak-coupling model.
This model is equivalent to that suggested by
Fabrizio and Gogolin.~\cite{rf:29}
It is shown that the two models provide us essentially the same expression
of the Josephson current
in both the high-temperature regime of $v_{\rm F}/L \ll T \ll \Delta$
and the low-temperature regime of $T \ll v_{\rm F}/L$.
We find that $j_{\rm c} \propto (1/L)^{2K_{\rho}^{-1} - 1}$
in the low-temperature regime.
Clearly, this result is inconsistent with
Fazio \textit{et al.}'s result,~\cite{rf:19, rf:20}
where the $L$-dependence is characterized by both $K_{\rho}$ and $K_{\sigma}$.
The incosistency is attributed to the difference in the treatment of
the potential induced by the NS contacts.
The basic assumption of Fazio \textit{et al.} is that
the superconductors do not influence the uniformity of the potential
along the 1D system,
while we assume that the low-energy property of an STLLS system
is described by the open-boundary fixed point.
Since the potential induced by the contacts becomes large
after renormalization even if it initially is very small,
Fazio \textit{et al.}'s result is unstable against
the renormalization unless the induced potential is completely negligible.
If the induced potential cannot be neglected,
the 1D system flows to the open-boundary fixed point.
That is, the 1D system is effectively disconnected at the NS contacts.
Fazio \textit{et al.}'s model is not appropriate in this situation,
and we expect that our weak-coupling model provides us correct results.

This paper is organized as follows.
In the next section, we present the interface-barrier model
on the basis of a bosonized description of an STLLS system.
The bosonization procedure based on Haldane's argument
is outlined in Appendix A.
In \S 3, we calculate the Josephson current in the interface-barrier model
using the instanton approximation.
The Josephson current in the weak-coupling model is studied in \S 4.
We employ the bosonized description of a TL liquid with open boundaries
given by Fabrizio and Gogolin,~\cite{rf:29}
and calculate the Josephson current within the lowest-order perturbation
with respect to the coupling between the TL liquid and the superconductors.
Section 5 is devoted to discussion.
The argument concerning to the interface-barrier model has been briefly
reported in ref.~\citen{rf:24}.
Some errors in ref.~\citen{rf:24} are corrected in this paper.
We set $\hbar = k_{\rm B} = 1$ throughout the paper.

\section{Interface-Barrier Model}

We consider a 1D electron system of length $L$ sandwiched between
superconductors (see Fig. 1(a)).
The superconducting order is characterized by the pair potential
$\Delta (x)$.
We assume that $\Delta (x)$ is given by~\cite{rf:28}
\begin{equation}
   \label{eq:Ddelta}
 \Delta (x) =
   \left\{ \begin{array}{cl}
      \Delta {\rm e}^{{\rm i} \chi_{1}} & \mbox{for $x \le 0$} \\
      0 & \mbox{for $0 < x < L $} \\
      \Delta {\rm e}^{{\rm i} \chi_{2}} & \mbox{for $L \le x$} ,
           \end{array}
   \right.
\end{equation}
where $\chi_{1}$ ($\chi_{2}$) is the macroscopic phase of
the left (right) superconsuctor.
We assume that the length $L$ is much longer than
the coherence length which is defined by
$\xi = v_{\rm F}/\Delta$ with the Fermi velocity $v_{\rm F}$.
A quasiparticle state in our system is described by
the Bogoliubov-de Gennes (BdG) equation.~\cite{rf:30}
Due to Andreev reflection at NS interfaces,
quasiparticle states depend on the phase differnce
$\chi = \chi_{2} - \chi_{1}$.
We introduce a set of eigenfunctions of the BdG equation.
The eigenfunctions are easily obtained
if we assume that perfect Andreev reflection (i.e., zero normal scattering)
is achieved at the NS interfaces.
In the perfect Andreev reflection case,
we obtain the low-energy eigenfunctions in the Nambu notation
\begin{align}
   \label{eq:eigen-f1}
 \left( \begin{array}{l} u_{q}(x) \\ v_{q}(x) \end{array} \right)
  & = \frac{1}{\sqrt{2 L}}
    \left( \begin{array}{l} 
            {\rm e}^{-{\rm i} \eta_{+}} {\rm e}^{{\rm i}q_{+}(x+\xi/2)} \\
            {\rm e}^{{\rm i} \eta_{+}} {\rm e}^{-{\rm i}q_{+}(x+\xi/2)}
          \end{array} \right)
       {\rm e}^{{\rm i}k_{\rm F}x} , \\
   \label{eq:eigen-f2}
 \left( \begin{array}{l} f_{q}(x) \\ g_{q}(x) \end{array} \right)
  & = \frac{1}{\sqrt{2 L}}
    \left( \begin{array}{l} 
            {\rm e}^{{\rm i} \eta_{-}} {\rm e}^{-{\rm i}q_{-}(x+\xi/2)} \\
            {\rm e}^{-{\rm i} \eta_{-}} {\rm e}^{{\rm i}q_{-}(x+\xi/2)}
          \end{array} \right)
       {\rm e}^{-{\rm i}k_{\rm F}x} ,
\end{align}
where $q = \pi ( n + 1/2)/(L + \xi)$ with $n = 0, \pm 1, \pm 2, \cdots$ and
\begin{align}
   \label{eq:q-n}
  q_{\pm} & = q \pm \frac{\chi}{2(L + \xi)} , \\
   \label{eq:phase-s}
  \eta_{\pm} & = \frac{\pi}{4} \mp \frac{\chi_{1}}{2} .
\end{align}
We have assumed that the Fermi wave number $k_{\rm F}$ is given by
$k_{\rm F} = \pi n_{0}/(L + \xi)$, where $n_{0}$ is an integer.
Let $\epsilon_{q+}$ and $\epsilon_{q-}$ be the linearized eigenenergy of
the state eq.~(\ref{eq:eigen-f1}) and that of the state
eq.~(\ref{eq:eigen-f2}), respectively.
They are given by
\begin{equation}
   \label{eq:eigen-e}
  \epsilon_{q\pm} = v_{\rm F} q_{\pm}
       + \frac{1}{2m} \left( \frac{\chi}{2(L + \xi)} \right)^{2}.
\end{equation}
In deriving eqs.~(\ref{eq:eigen-f1}) and (\ref{eq:eigen-f2}),
we have used the following approximation
\begin{equation}
  \frac{\epsilon_{q\pm} + {\rm i} \sqrt{\Delta^{2} - \epsilon_{q\pm}^{2}}}
       {\Delta}
  = \exp {\rm i}
         \left( \frac{\pi}{2} - \frac{\epsilon_{q\pm}}{\Delta}
         \right) ,
\end{equation}
which is justified when $\epsilon_{q\pm} \ll \Delta$.

We assume that any quasiparticle state in our system is described
by eqs.~(\ref{eq:eigen-f1}) or (\ref{eq:eigen-f2}).
This is the key approximation allowing us to perform
the bosonization procedure.
Clearly, this cannot be justified for $q$ of the order of,
or greater than, $\xi^{-1}$.
However, high-energy states compared with $\Delta$ play
only a minor role in low-energy properties, which are of interest to us.
To study Josephson effect,
we must take account of the zero modes~\cite{rf:17}
as well as the nonzero modes.
To do so, we introduce the winding-number operators $J$ and $M$,
and the zero-mode operators $\varphi_{\rho}$ and $\varphi_{\sigma}$,
which satisfy
$[J, \varphi_{\rho}] = [M, \varphi_{\sigma}]= 2 {\rm i}$
and $[J, \varphi_{\sigma}] = [M, \varphi_{\rho}]= 0$.
The operator $J$ ($M$) is related to the excess charge (spin)
accumulated in the 1D system.
The nonzero modes are described by the boson operators
$\alpha_{q}$, $\alpha_{q}^{\dagger}$, $\beta_{q}$ and $\beta_{q}^{\dagger}$.
The bosonization procedure is outlined in Appendix A.

In terms of the operators introduced above,
the bosonized Hamiltonian is given by
\begin{align}
   \label{eq:Hamil-00}
 H & = \frac{\pi}{4L} \left( 
          v_{\rho} K_{\rho} \left( J + \frac{\chi}{\pi} \right)^{2}
          + \frac{v_{\sigma}}{K_{\sigma}} M^{2} \right)
               \nonumber \\
   & \hspace{10mm}
        + \sum_{q>0} \left( v_{\rho} q a_{q}^{\dag} a_{q}
                          + v_{\sigma} q b_{q}^{\dag} b_{q} \right) ,
\end{align}
where $v_{\rho}$ ($v_{\sigma}$) is the velocity of charge (spin) excitation
and  $q = \pi (n+1/2)/(L+\xi)$ with $n = 0, 1, 2, \cdots$.  
In eq.~(\ref{eq:Hamil-00}),
$K_{\rho}$ and $K_{\sigma}$ are the correlation exponents, which
characterize the interaction strengths.
For example, $K_{\rho} < 1$ and $K_{\sigma} = 1$ corresponds to
the spin-independent repulsive interaction case,
and the system is reduced to the noninteracting electron gas when
$K_{\rho} = K_{\sigma} = 1$.
It is important to note that eigenvalues of $J + M$ are limited to be
even integers, as shown in Appendix A.

We express the electron-field operator as $\psi_{s}(x)$,
where $s = \pm$ denotes spin index,
and decompose it as
\begin{equation}
   \label{eq:fop-de}
   \psi_{s}(x) = \psi_{1s}(x) + \psi_{2s}(x),
\end{equation}
where $\psi_{1s}$ and $\psi_{2s}$ represent
the right and left movers, respectively.
To express $\psi_{1s}$ and $\psi_{2s}$,
we define the phase fields:~\cite{rf:31}
\begin{align}
   \label{eq:Theta-+}
  \Theta_{+}(x) & = \varphi_{\rho} + \theta_{+}(x) , \\
   \label{eq:Theta--}
  \Theta_{-}(x) & = \frac{\pi}{L} \left( J + \frac{\chi}{\pi} \right)
                            \left( x + \frac{\xi}{2} \right)
                        + \theta_{-}(x) , \\
   \label{eq:Phi-+}
  \Phi_{+}(x) & = \frac{\pi}{L} M 
                           \left( x + L + \frac{\xi}{2} \right)
                        + \phi_{+}(x) , \\
   \label{eq:Phi--}
  \Phi_{-}(x) & = \varphi_{\sigma} + \phi_{-}(x) .
\end{align}
The nonzero-mode components $\theta_{\pm}$ and $\phi_{\pm}$,
which describe low-energy fluctuations around the zero modes,
are expressed in terms of the boson operators as
\begin{align}
   \label{eq:f-theta-+}
  \theta_{+}(x) & =
         {\rm i} \sqrt{K_{\rho}} \sum_{q>0} \sqrt{\frac{2\pi}{q L}}
              {\rm e}^{-\alpha q/2} \cos q \left( x + \frac{\xi}{2} \right)
              \left( a_{q}^{\dag} - a_{q}  \right) , \\
   \label{eq:f-theta--}
  \theta_{-}(x) & =
         \frac{1}{\sqrt{K_{\rho}}} \sum_{q>0} \sqrt{\frac{2\pi}{q L}}
              {\rm e}^{-\alpha q/2} \sin q \left( x + \frac{\xi}{2} \right)
              \left( a_{q}^{\dag} + a_{q}  \right) , \\
   \label{eq:f-phi-+}
  \phi_{+}(x) & =
         \sqrt{K_{\sigma}} \sum_{q>0} \sqrt{\frac{2\pi}{q L}}
              {\rm e}^{-\alpha q/2} \sin q \left( x + \frac{\xi}{2} \right)
              \left( b_{q}^{\dag} + b_{q}  \right) , \\
   \label{eq:f-phi--}
  \phi_{-}(x) & =
         \frac{\rm i}{\sqrt{K_{\rho}}} \sum_{q>0} \sqrt{\frac{2\pi}{q L}}
              {\rm e}^{-\alpha q/2} \cos q \left( x + \frac{\xi}{2} \right)
              \left( b_{q}^{\dag} - b_{q}  \right) ,
\end{align}
where $\alpha$ is a positive infinitesimal.
The phase field $\Theta_{+}$ is regarded as the phase of
the charge density wave,
while $\Theta_{-}$ corresponds to the Josephson phase.
The phase fields $\Phi_{+}$ and $\Phi_{-}$ are related to
the spin degrees of freedom.
Using the phase fields, we can express $\psi_{1s}$ and $\psi_{2s}$
as~\cite{rf:32}
\begin{align}
   \label{eq:psi-1}
   \psi_{1s}(x) & = \frac{1}{\sqrt{2 \pi \alpha}}
                  \exp \bigg( {\rm i} k_{\rm F} x + \frac{\rm i}{2} 
               \bigl( \Theta_{+}(x) + \Theta_{-}(x)
                  \nonumber \\
                & \hspace{25mm}
                        + s \Phi_{+}(x) + s \Phi_{-}(x) \bigr) \bigg) , \\
   \label{eq:psi-2}
   \psi_{2s}(x) & = \frac{1}{\sqrt{2 \pi \alpha}}
                  \exp \bigg( - {\rm i} k_{\rm F} x + \frac{\rm i}{2} 
               \bigl( -\Theta_{+}(x) + \Theta_{-}(x)
                  \nonumber \\
                & \hspace{25mm}
                        - s \Phi_{+}(x) + s \Phi_{-}(x) \bigr) \bigg) .
\end{align}

Now we put a barrier at each NS interface.
For simplicity, we assume that both the barriers have the same strength.
Then, the corresponding Hamiltonian is
\begin{align}
   \label{eq:Hamil-b}
  H^{\rm B} & = - V_{0} \sum_{s} \biggl(
        \left( \psi_{1s}^{\dag}(0) \psi_{2s}(0) + {\rm h.c.} \right)
                  \nonumber \\
            & \hspace{20mm}
      + \left( \psi_{1s}^{\dag}(L) \psi_{2s}(L) + {\rm h.c.} \right)
                             \biggr) .
\end{align}
To express $H^{\rm B}$, it is convenient to introduce the new
phase variables:~\cite{rf:8,rf:9}
\begin{align}
   \label{eq:bar-t}
  \bar{\theta} & = \frac{1}{2} \left(
              \theta_{+}(L) + \theta_{+}(0) \right) , \\
   \label{eq:tilde-t}
  \tilde{\theta} & = \left(
              \theta_{+}(L) - \theta_{+}(0) \right) , \\
   \label{eq:bar-p}
  \bar{\phi} & = \frac{1}{2} \left(
              \phi_{+}(L) + \phi_{+}(0) \right) , \\
   \label{eq:tilde-p}
  \tilde{\phi} & = \left(
              \phi_{+}(L) - \phi_{+}(0) \right) .
\end{align}
The time-derivative of $\bar{\theta}$ ($\bar{\phi}$) is related to
the charge (spin) current,
while $\tilde{\theta}$ ($\tilde{\phi}$) is related to
the excess charge (spin) accumulated in the 1D system.
In terms of them, we find that
\begin{align}
   \label{eq:Hamil-b2}
  H^{\rm B}
 & = - \frac{4 V_{0}}{\pi \alpha} \Bigg[
         \cos \left( \varphi_{\rho} + \bar{\theta} + k_{\rm F}L \right)
         \cos \left( \frac{\tilde{\theta}}{2} + k_{\rm F}L \right)
                   \nonumber \\
 & \hspace{20mm} \times
         \cos \left( \bar{\phi} + \frac{3}{2} \pi M \right)
         \cos \left( \frac{\tilde{\phi}}{2} + \pi M \right)
                   \nonumber \\
 & \hspace{14mm}
       + \sin \left( \varphi_{\rho} + \bar{\theta} + k_{\rm F}L \right)
         \sin \left( \frac{\tilde{\theta}}{2} + k_{\rm F}L \right)
                   \nonumber \\
 & \hspace{20mm} \times
         \sin \left( \bar{\phi} + \frac{3}{2} \pi M \right)
         \sin \left( \frac{\tilde{\phi}}{2} + \pi M \right)
                                           \Bigg]  ,
\end{align}
where we approximated as
$(3 \pi M/2)(1 + \xi / (3L) ) \approx 3 \pi M/2$.

\section{Josephson Current in the Interface-Barrier Model}

In this section, we consider the interface-barrier model in the limit
where the barrier potential is very strong.
The total Hamiltonian of our system is $H_{\rm total} = H + H^{\rm B}$.
We calculate the partition function $Z$ in terms of
an imaginary-time path integral.
The dc Josephson current is obtained through the well-known relation
\begin{equation}
   \label{eq:j-from-Z}
 j(\chi) = -2eT \frac{\partial \ln Z}{\partial \chi} .
\end{equation}
Since $M$ commutes with $H^{\rm B}$, then $M$ remains a conserved quantity
despite the presence of the barriers at the NS interfaces.
By contrast, $J$ is not a good quantum number when $V_{0} \neq 0$.
The partition function is expressed as
\begin{align}
   \label{eq:pf-1}
  Z
 & = \sum_{M = - \infty \atop (M = {\rm even})}^{\infty}
         \exp \left(-\beta \frac{\pi v_{\sigma} M^{2}}{4K_{\sigma}L}
              \right) Z_{+}
                  \nonumber \\
 & \hspace{15mm}
    + \sum_{M = - \infty \atop (M = {\rm odd})}^{\infty}
         \exp \left(-\beta \frac{\pi v_{\sigma} M^{2}}{4K_{\sigma}L}
              \right) Z_{-} ,
\end{align}
where
\begin{align}
   \label{eq:pf-M}
  Z_{\pm}
 & = \sum_{m = -\infty}^{\infty}
            \int_{\Delta \varphi_{\rho} = 2\pi m}
                \mathcal{D} \varphi_{\rho}
            \int \prod_{q>0}
                 \mathcal{D}a_{q}^{\dagger} \mathcal{D}a_{q}
                 \mathcal{D}b_{q}^{\dagger} \mathcal{D}b_{q}
                  \nonumber \\
 & \hspace{5mm} \times
            \exp \left( {\rm i}\int_{0}^{\beta} {\rm d} \tau 
                         \frac{\chi + \zeta_{\pm}}{2 \pi}\partial_{\tau}
                                  \varphi_{\rho}
                       - S^{\rm Z} - S^{\rm NZ} - S^{\rm B} \right) ,
\end{align}
with $\zeta_{+} = 0$ and $\zeta_{-} = \pi$, and
\begin{align}
   \label{eq:S-0-Z}
  S^{\rm Z} & = \int_{0}^{\beta} {\rm d} \tau
                          \frac{L}{4 \pi v_{\rho} K_{\rho}}
                          (\partial_{\tau} \varphi_{\rho})^{2} , \\
   \label{eq:S-0-NZ}
  S^{\rm NZ} & = \int_{0}^{\beta} {\rm d} \tau \sum_{q>0} \big(
                   a_{q}^{\dagger} \partial_{\tau} a_{q}
                       + v_{\rho} q a_{q}^{\dagger} a_{q}
                   \nonumber \\
             & \hspace{25mm}
                 + b_{q}^{\dagger} \partial_{\tau} b_{q}
                       + v_{\sigma} q b_{q}^{\dagger} b_{q} \big) , \\
   \label{eq:S-B}
  S^{\rm B} & = \int_{0}^{\beta} {\rm d} \tau H^{\rm B} (\tau ) .
\end{align}
In eq.~(\ref{eq:pf-M}), the integration over $\varphi_{\rho}$ is carried out
under the boundary condition of
$\varphi_{\rho} (\beta) - \varphi_{\rho} (0) = 2\pi m$.
This boundary condition is imposed to take
the discrete nature of $J$ into account.
Furthermore, we used $\zeta_{\pm}$ to maintain the constraint
$J + M = \mbox{even}$.~\cite{rf:33}

Before considering the strong-barrier limit, which is of interest to us,
we treat the case of $V_{0} \to 0$ to examine
the validity of eq.~(\ref{eq:pf-1}).
When $S^{\rm B} = 0$, the partition function factrizes as
$Z = \bar{Z(\chi)} \tilde{Z}$,
where $\bar{Z(\chi)}$ and $\tilde{Z}$ are the contribution
from the zero modes and that from the nonzero modes, respectively.
The former contribution $\bar{Z}$ is given by
\begin{align}
   \label{eq:bar-pf}
 \bar{Z} 
  & = \sum_{M = - \infty \atop (M = {\rm even})}^{\infty}
           \exp \left(-\beta \frac{\pi v_{\sigma} M^{2}}{4K_{\sigma}L}
                \right) \bar{Z}_{+}
                  \nonumber \\
 & \hspace{20mm}
         + \sum_{M = - \infty \atop (M = {\rm odd})}^{\infty}
           \exp \left(-\beta \frac{\pi v_{\sigma} M^{2}}{4K_{\sigma}L}
                \right) \bar{Z}_{-} ,
\end{align}
where
\begin{align}
   \label{eq:bar-pf-pm}
   \bar{Z}_{\pm}
  & = \sum_{m = - \infty}^{\infty} \int_{\Delta \varphi_{\rho} = 2\pi m}
                   \mathcal{D} \varphi_{\rho}
                  \nonumber \\
 & \hspace{15mm} \times
           \exp \left( {\rm i}\int_{0}^{\beta} {\rm d} \tau 
               \frac{\chi + \zeta_{\pm}}{2 \pi}\partial_{\tau} \varphi_{\rho}
                          - S^{\rm Z} \right) .
\end{align}
The latter contribution $\tilde{Z}$, which does not depend on $\chi$,
is irrelevant for the Josephson effect.
We decompose $\varphi_{\rho}$ as
$\varphi_{\rho}(\tau) = \varphi_{\rho 0} +2 \pi m \tau /\beta
+ \tilde{\varphi}_{\rho}(\tau)$
with $\tilde{\varphi}_{\rho}(\beta) = \tilde{\varphi}_{\rho}(0)$.
Substitution of this into eq.~(\ref{eq:bar-pf-pm}) yields
\begin{equation}
   \label{eq:bar-pf2}
  \bar{Z}_{\pm} =
            \sum_{m = - \infty}^{\infty}
            \exp \Bigl( - \frac{\pi L}{\beta v_{\rho} K_{\rho}} m^{2}
                         + {\rm i}m (\chi + \zeta_{\pm}) \Bigr) .
\end{equation}
Applying Poisson's summation theorem to eq.~(\ref{eq:bar-pf2})
and then substituting the resulting expression into eq.~(\ref{eq:bar-pf}),
we obtain the correct partition function
in the absence of the barriers,~\cite{rf:21}
\begin{align}
   \label{eq:bar-pf3}
\bar{Z} & \propto \sum_{J, M \atop (J + M = {\rm even})}
          \exp \bigg( - \beta \frac{\pi K_{\rho} v_{\rho}}{4 L}
                       \left( J + \frac{\chi}{\pi}\right)^{2}
                  \nonumber \\
 & \hspace{35mm}
                      - \beta \frac{\pi v_{\sigma}}{4 K_{\sigma }L} M^{2}
               \bigg) ,
\end{align}
where the summation over $J$ and $M$ is carried out
under the condition of $J + M = \mbox{even}$.

Now we turn to the strong-barrier limit, where $H^{\rm B}$ is regarded as
a strong pinning potential for the phase fields
$\bar{\theta}$, $\tilde{\theta}$, $\bar{\phi}$ and $\tilde{\phi}$.
It is convenient to introduce the effective action which satisfies
\begin{align}
   \label{eq:def-S-eff-NZ}
 & \frac{ \int \prod_{q>0} \mathcal{D}a_{q}^{\dagger} \mathcal{D}a_{q}
                   \mathcal{D}b_{q}^{\dagger} \mathcal{D}b_{q}
                   \exp \left( - S^{\rm NZ} - S^{\rm B} \right) }
       { \int \prod_{q>0} \mathcal{D}a_{q}^{\dagger} \mathcal{D}a_{q}
                   \mathcal{D}b_{q}^{\dagger} \mathcal{D}b_{q}
                   \exp \left( - S^{\rm NZ} \right) }
                   \nonumber \\
 & \hspace{10mm} =
  \frac{ \int \mathcal{D} \bar{\theta} \mathcal{D} \tilde{\theta}
              \mathcal{D} \bar{\phi} \mathcal{D} \tilde{\phi}
                   \exp \left( - S_{\rm eff}^{\rm NZ} - S^{\rm B} \right) }
       { \int \mathcal{D} \bar{\theta} \mathcal{D} \tilde{\theta}
              \mathcal{D} \bar{\phi} \mathcal{D} \tilde{\phi}
                   \exp \left( - S_{\rm eff}^{\rm NZ} \right) } .
\end{align}
This action is given by
\begin{align}
   \label{eq:S-eff-NZ}
S_{\rm eff}^{\rm NZ}
 &     = \frac{1}{2 \pi K_{\rho} \beta} \sum_{\omega}
            \bar{J}_{\rho}^{-1}(\omega)
                 \bar{\theta}(\omega) \bar{\theta}(-\omega)
                  \nonumber \\
 & \hspace{5mm}
       + \frac{1}{8 \pi K_{\rho} \beta} \sum_{\omega}
            \tilde{J}_{\rho}^{-1}(\omega)
                 \tilde{\theta}(\omega) \tilde{\theta}(-\omega)
                  \nonumber \\
 & \hspace{5mm}
       + \frac{1}{2 \pi K_{\sigma} \beta} \sum_{\omega}
            \bar{J}_{\sigma}^{-1}(\omega)
                 \bar{\phi}(\omega) \bar{\phi}(-\omega)
                  \nonumber \\
 & \hspace{5mm}
       + \frac{1}{8 \pi K_{\sigma} \beta} \sum_{\omega}
            \tilde{J}_{\sigma}^{-1}(\omega)
                 \tilde{\phi}(\omega) \tilde{\phi}(-\omega) ,
\end{align}
where
\begin{align}
   \label{eq:bar-J-rho}
 \bar{J}_{\rho}(\omega)
 & = \frac{1}{2 \omega \sinh \left( \frac{L \omega}{v_{\rho}} \right)}
     \Bigg\{  \cosh \left( \frac{L \omega}{v_{\rho}} \right)
         + \cosh \left( \frac{(L-\xi) \omega}{v_{\rho}} \right)
                   \nonumber \\
 & \hspace{20mm}
         + \cosh \left( \frac{\xi \omega}{v_{\rho}} \right) +1 \Bigg\}
            -\frac{2v_{\rho}}{L \omega^{2}} , \\
   \label{eq:tilde-J-rho}
 \tilde{J}_{\rho}(\omega)
 & = \frac{1}{2 \omega \sinh \left( \frac{L \omega}{v_{\rho}} \right)}
     \Bigg\{  \cosh \left( \frac{L \omega}{v_{\rho}} \right)
         + \cosh \left( \frac{(L-\xi) \omega}{v_{\rho}} \right)
                    \nonumber \\
 & \hspace{20mm}
         - \cosh \left( \frac{\xi \omega}{v_{\rho}} \right) -1 \Bigg\} , \\
   \label{eq:bar-J-sigma}
 \bar{J}_{\sigma}(\omega)
 & = \frac{1}{2 \omega \sinh \left( \frac{L \omega}{v_{\sigma}} \right)}
     \Bigg\{  \cosh \left( \frac{L \omega}{v_{\sigma}} \right)
         - \cosh \left( \frac{(L-\xi) \omega}{v_{\sigma}} \right)
                   \nonumber \\
 & \hspace{20mm}
         - \cosh \left( \frac{\xi \omega}{v_{\sigma}} \right) +1 \Bigg\} , \\
   \label{eq:tilde-J-sigma}
 \tilde{J}_{\sigma}(\omega)
 & = \frac{1}{2 \omega \sinh \left( \frac{L \omega}{v_{\sigma}} \right)}
     \Bigg\{  \cosh \left( \frac{L \omega}{v_{\sigma}} \right)
         - \cosh \left( \frac{(L-\xi) \omega}{v_{\sigma}} \right)
                   \nonumber \\
 & \hspace{20mm}
         + \cosh \left( \frac{\xi \omega}{v_{\sigma}} \right) -1 \Bigg\} .
\end{align}
The derivation of the effective action is outlined in Appendix B.
From eqs.~(\ref{eq:bar-J-sigma}) and (\ref{eq:tilde-J-sigma}),
we see that $\bar{J}_{\sigma}^{-1}(\omega)$ and
$\tilde{J}_{\sigma}^{-1}(\omega)$ are of the order of $\Delta$
in the limit of $\omega \to 0$.
This indicates that $\bar{\phi}$ and $\tilde{\phi}$ have a mass gap
of the order of $\Delta$
and thereby their low-energy fluctuations are strongly suppressed.
We thus see that $\bar{\phi}$ and $\tilde{\phi}$ are pinned
by $H^{\rm B}$ when $T \ll \Delta$.
This is an important feature of the phase variables in our problem.
We can understand the feature by noting that
only a spin-singlet pair of two electrons can transfer the NS interfaces
due to Andreev reflection.
Detailed discussion on the characteristic features
of $S_{\rm eff}^{\rm NZ}$ is given in \S 5.

Since $\bar{\phi}$ and $\tilde{\phi}$ are pinned by the barrier potential,
we can set $\bar{\phi} = \tilde{\phi} = 0$ without loss of generality.
Furthermore, we note that $M$ plays only a minor role in changing
the location of the potential minima,
at which $\varphi + \bar{\theta}$ and $\tilde{\theta}$ are pinned.
Consequently, $S^{\rm B}$ can be simplified to
\begin{equation}
   \label{eq:S-B2}
  S^{\rm B} = - \frac{4 V_{0}}{\pi \alpha} \int_{0}^{\beta} {\rm d} \tau
              \cos (\varphi_{\rho} + \bar{\theta})
              \cos \frac{\tilde{\theta}}{2} ,
\end{equation}
where $k_{\rm F}L = 0$ (mod $\pi$) is assumed for simplicity.
Since $\bar{\phi}$ and $\tilde{\phi}$ are irrelevant for our argument,
$S_{\rm eff}^{\rm NZ}$ is reduced to
\begin{align}
   \label{eq:S-eff-NZ2}
S_{\rm eff}^{\rm NZ}
   &  = \frac{1}{2 \pi K_{\rho} \beta} \sum_{\omega}
            \bar{J}_{\rho}^{-1}(\omega)
                 \bar{\theta}(\omega) \bar{\theta}(-\omega)
                   \nonumber \\
   & \hspace{7mm}
       + \frac{1}{8 \pi K_{\rho} \beta} \sum_{\omega}
            \tilde{J}_{\rho}^{-1}(\omega)
                 \tilde{\theta}(\omega) \tilde{\theta}(-\omega) .
\end{align}
It should be noted here that the high-frequency cutoff $\Lambda / 2$
($2 \Lambda$) must be introduced for the summation in the first (second) term
to avoid ultraviolet divergence ($\Lambda$ is of the order of $\Delta$).
To do so, we introduce an additional action which serves
as the high-frequency cutoff for $S_{\rm eff}^{\rm NZ}$,~\cite{rf:10}
\begin{equation}
   \label{eq:S-Lambda}
 S^{\Lambda} = \int_{0}^{\beta} {\rm d} \tau \left(
          \frac{\bar M}{2}
                       \left( \partial_{\tau} \bar{\theta} \right)^{2}
        + \frac{\tilde M}{2} 
                       \left( \partial_{\tau} \tilde{\theta} \right)^{2}
                                 \right) ,
\end{equation}
where $\bar{M} \approx 2 / \Lambda$ and $\tilde{M} \approx 1/ (2 \Lambda)$.
After these treatments, eq.~(\ref{eq:pf-M}) is reduced to
\begin{align}
   \label{eq:pf-M2}
  Z_{\pm}
 & = \sum_{m = -\infty}^{\infty}
     {\rm e}^{{\rm i} m (\chi + \zeta_{\pm})}
     \int_{\Delta \varphi_{\rho} = 2\pi m} \mathcal{D} \varphi_{\rho}
     \int \mathcal{D} \bar{\theta} \mathcal{D} \tilde{\theta}
                   \nonumber \\
 & \hspace{10mm} \times
          \exp \left( - S^{\rm Z} - S_{\rm eff}^{\rm NZ}
                       - S^{\Lambda}- S^{\rm B} \right) .
\end{align}

In the strong-barrier limit, the change in $\varphi_{\rho}$ with $m \neq 0$
occurs in collaboration with instanton tunneling of
$\bar{\theta}$ and $\tilde{\theta}$ from a potential minimum
$(\varphi_{\rho}+\bar{\theta}, \tilde{\theta}) = (k \pi, 2l \pi)$
to an adjacent minimum
$(\varphi_{\rho}+\bar{\theta}, \tilde{\theta})
= ((k \pm 1) \pi, (2l \pm 2) \pi)$, where $k$ and $l$ are integers.
The matrix element $\gamma$ for this tunneling process is very small,
so that we are allowed to consider only the second-order processes
with respect to $\gamma$.
Thus, these lowest-order tunneling processes correspond to
the change in $\varphi_{\rho}$ with $m = \pm 1$.
Within this lowest-oredr approximation,
$Z_{\pm}$ is simplified to
\begin{equation}
   \label{eq:pf-M3}
  Z_{\pm} \propto 1 \pm 2 \cos \chi \cdot Z_{1}/Z_{0} ,
\end{equation}
where
\begin{align}
   \label{eq:m-pf-j}
  Z_{m}
  & = \int_{\Delta \varphi_{\rho} = 2\pi m} \mathcal{D} \varphi_{\rho}
      \int \mathcal{D} \bar{\theta} \mathcal{D} \tilde{\theta}
                   \nonumber \\
 & \hspace{10mm} \times
      \exp \left( - S^{\rm Z} - S_{\rm eff}^{\rm NZ}
                       - S^{\Lambda}- S^{\rm B} \right) . 
\end{align}
To calculate $Z_{1}$, we first determine the stationary path
of $S_{\rm st} \equiv S^{\rm Z} + S^{\Lambda} + S^{\rm B}$
under the condition of
$\varphi_{\rho}(\beta) = \varphi_{\rho}(0) + 2 \pi$,
and then incorporate the influence of $S_{\rm eff}^{\rm NZ}$,
which plays the role of a dissipative environment,
by integrating out the low-energy fluctuations around the stationary path.
We introduce the new variables:
\begin{align}
   \label{eq:theta-r1}
   \theta_{r} & = \varphi_{\rho} + \bar{\theta} , \\
   \label{eq:rheta-r2}
   \theta_{R} & = \frac{1}{m_{L}+\bar{M}}
                \left( m_{L} \varphi_{\rho} - \bar{M} \bar{\theta} \right) ,
\end{align}
where $m_{L} = L/(2\pi v_{\rho}K_{\rho})$.
Note that $\theta_{r}$ ($\theta_{R}$) is the relative (center-of-mass)
coordinate with respect to $\varphi_{\rho}$ and $- \bar \theta$.
Then, $S_{\rm st}$ becomes
\begin{align}
   \label{eq:S-st}
S_{\rm st}
 & = \int_{0}^{\beta} {\rm d} \tau
    \Bigg(
        \frac{M_{r}}{2} ( \partial_{\tau}\theta_{r} )^{2}
      + \frac{M_{R}}{2} (\partial_{\tau}\theta_{R} )^{2}
      + \frac{\tilde M}{2} (\partial_{\tau}\tilde{\theta} )^{2}
                   \nonumber \\
 & \hspace{30mm}
      - \frac{4V_{0}}{\pi \alpha} \cos\theta_{r} \cos \frac{\tilde \theta}{2}
                                          \Bigg) ,
\end{align}
where $M_{r} = m_{L} \bar{M}/(m_{L}+\bar{M})$ and $M_{R}=m_{L}+\bar{M}$.
The stationary path is determind under the boundary condition of
$\theta_{R}(\beta) = \theta_{R}(0) + 2 \pi m_{L}/(m_{L}+\bar{M})$
and $\theta_{r}(\beta) = \theta_{r}(0) + 2 \pi$.
Noting that $m_{L} \gg \bar{M}$, we find that
$\theta_{R} \approx \varphi_{\rho}$.
Thus, we approximately obtain the stationary path as
\begin{align}
   \label{eq:varphi-st0}
  \theta_{R}^{\rm st} & \approx
  \varphi_{\rho}^{\rm st} = \frac{2 \pi}{\beta} \tau , \\
   \label{eq:bar-theta-st}
  \theta_{r}^{\rm st} & = \frac{1}{2} I(\tau - \tau_{1})
                           + \frac{1}{2} I(\tau - \tau_{2}) , \\
   \label{eq:tilde-theta-st}
  \tilde{\theta}^{\rm st} & = I(\tau - \tau_{1}) - I(\tau - \tau_{2}) ,
\end{align}
where $I(\tau - \tau_{i})$ describes one instanton at $\tau = \tau_{i}$
and satisfies $I(- \infty) = 0$ and $I(\infty) = 2 \pi$.
Then, we find that
\begin{equation}
  \bar{\theta}^{\rm st} =  \frac{1}{2} I(\tau - \tau_{1})
                         + \frac{1}{2} I(\tau - \tau_{2})
                           - \frac{2 \pi}{\beta} \tau .
\end{equation}

We calculate $Z_{1}$ taking account of
the low-energy fluctuations around the stationary path.
Due to the barrier potential,
$\tilde \theta$ and $\varphi_{\rho} + \bar{\theta}$ are strongly pinned.
Thus, we can neglect the fluctuations of $\tilde \theta$ around
$\tilde{\theta}^{\rm st}$.
In contrast, $\varphi_{\rho}$ and $\bar{\theta}$ can fluctuate
under the condition that
$\varphi_{\rho} + \bar{\theta}$ is strongly pinned.
This means that they are expressed as
\begin{align}
   \label{eq:varphi-dec}
  \varphi_{\rho} & = \varphi_{\rho}^{\rm st} + \phi , \\
   \label{eq:var-theta-dec}
  \bar{\theta} & = \bar{\theta}^{\rm st} - \phi ,
\end{align}
where $\phi$ represents the low-energy fluctuations.
We approximate that $I(\tau ) = 2 \pi \vartheta (\tau )$,
where $\vartheta (\tau )$ is the step function.
The Fourier transform of $\bar{\theta}^{\rm st}$
and that of $\tilde{\theta}^{\rm st}$ are expressed as
\begin{align}
   \label{eq:bar-theta-st2}
  \bar{\theta}^{\rm st} (\omega)
  & =    \frac{{\rm i}\pi}{\omega} {\rm e}^{{\rm i}\omega \tau_{1}}
       + \frac{{\rm i}\pi}{\omega} {\rm e}^{{\rm i}\omega \tau_{2}} , \\
   \label{eq:tilde-theta-st2}
  \tilde{\theta}^{\rm st} (\omega)
  & =    \frac{{\rm i}2\pi}{\omega} {\rm e}^{{\rm i}\omega \tau_{1}}
       - \frac{{\rm i}2\pi}{\omega} {\rm e}^{{\rm i}\omega \tau_{2}} .
\end{align}
Substituting eqs.~(\ref{eq:bar-theta-st2}) and (\ref{eq:tilde-theta-st2})
into eq.~(\ref{eq:m-pf-j}),
we find that
\begin{equation}
   \label{eq:m-pf-1}
  \frac{Z_{1}}{Z_{0}}
     = \gamma^{2} \int_{0}^{\beta} {\rm d}\tau_{1}{\rm d}\tau_{2}
            \int \mathcal{D} \phi \, {\rm e}^{ - S_{\rm ins}} ,
\end{equation}
where
\begin{align}
   \label{eq:S-ins}
 S_{\rm ins}
 & =  \frac{\pi L}{v_{\rho} K_{\rho} \beta}
    + \frac{1}{\beta} \sum_{\omega > 0} 
        \frac{L \omega^{2}}{2\pi v_{\rho} K_{\rho} }
        \phi (\omega) \phi(-\omega) 
               \nonumber \\
 & \hspace{5mm}
     + \frac{1}{\pi K_{\rho} \beta} \sum_{\omega > 0}
        \bar{J}_{\rho}^{-1}(\omega)
        \left( \bar{\theta}^{\rm st}(\omega) - \phi(\omega) \right)
                   \nonumber \\
               & \hspace{35mm} \times
        \left( \bar{\theta}^{\rm st}(- \omega) - \phi(- \omega) \right)
               \nonumber \\
 & \hspace{5mm}
     + \frac{1}{4\pi K_{\rho} \beta} \sum_{\omega > 0}
        \tilde{J}_{\rho}^{-1}(\omega)
        \tilde{\theta}^{\rm st}(\omega) \tilde{\theta}^{\rm st}(- \omega) .
\end{align}
Integrating out $\phi$, we obtain
\begin{equation}
   \label{eq:m-pf-2}
  \frac{Z_{1}}{Z_{0}}
   = \gamma^{2} \exp \left( -\frac{\pi L}{K_{\rho} v_{\rho}} T \right)
        \int_{0}^{\beta} {\rm d}\tau_{1}{\rm d}\tau_{2} \, 
             {\rm e}^{-\tilde{S}_{\rm ins}} ,
\end{equation}
where
\begin{align}
   \label{eq:S-ins2}
 \tilde{S}_{\rm ins}
 & = \frac{4 \pi}{K_{\rho} \beta} \sum_{\omega > 0}
     \frac{1}{ \frac{2v_{\rho}}{L}
                  + \omega^{2} \bar{J}_{\rho}(\omega) }
                   \nonumber \\
 & \hspace{3mm}
      + \frac{2 \pi}{K_{\rho} \beta} \sum_{\omega > 0}
         \left(
          \frac{1}{ \omega^{2} \tilde{J}_{\rho}(\omega) }
        - \frac{1}{ \frac{2v_{\rho}}{L} + \omega^{2} \bar{J}_{\rho}(\omega) }
         \right)
                   \nonumber \\
 & \hspace{23mm} \times
        \bigl(1 - \cos \omega \left(\tau_{1}-\tau_{2}\right) \bigr) .
\end{align}
We substitute eqs.~(\ref{eq:bar-J-rho}) and (\ref{eq:tilde-J-rho})
into eq.~(\ref{eq:S-ins2}).
Noting that $L \gg \xi$, we obtain
\begin{align}
   \label{eq:S-ins3}
 \tilde{S}_{\rm ins}
 & = \frac{4 \pi}{K_{\rho} \beta} \sum_{\Lambda > \omega > 0}
     \frac{1}{\omega} \tanh \left( \frac{L \omega}{2v_{\rho}} \right)
                   \nonumber \\
 & \hspace{3mm}
   + \frac{4 \pi}{K_{\rho} \beta} \sum_{\omega > 0}
     \frac{1}{\omega} {\rm cosech} \left( \frac{L \omega}{v_{\rho}}
                                   \right)
     \bigl(1 - \cos \omega \left(\tau_{1}-\tau_{2}\right) \bigr) ,
\end{align}
where we have introduced the high-frequency cutoff $\Lambda$ in the first term.

We first consider the high-temperature regime of
$v_{\rm F}/L \ll T \ll \Delta$.
Noting that
\begin{align}
   \label{eq:sum-M1}
 \sum_{M = - \infty \atop (M = {\rm even})}^{\infty}
    \exp \left( - \beta \frac{\pi v_{\sigma}}{4K_{\sigma}L} M^{2} \right)
  & \propto
      1 + 2 \exp \left(-\frac{\pi K_{\sigma}L}{v_{\sigma}}T \right) ,
            \\
   \label{eq:sum-M2}
 \sum_{M = - \infty \atop (M = {\rm odd})}^{\infty}
    \exp \left( -\beta \frac{\pi v_{\sigma}}{4K_{\sigma}L} M^{2} \right)
  & \propto
      1 - 2 \exp \left(-\frac{\pi K_{\sigma}L}{v_{\sigma}}T \right) ,
\end{align}
the partition function is expressed as
\begin{align}
   \label{eq:pf-high}
 Z & \propto 2 + 8 \cos \chi \cdot \gamma^{2}
       \exp \left( - \pi \left( \frac{L}{v_{\rho} K_{\rho}}
               +\frac{K_{\sigma} L}{v_{\sigma}} \right) T \right) 
                   \nonumber \\
 & \hspace{30mm} \times
       \int_{0}^{\beta} {\rm d}\tau_{1}{\rm d}\tau_{2} \, 
       {\rm e}^{- \tilde{S}_{\rm ins}} .
\end{align}
In the high-temperature regime, $\tilde{S}_{\rm ins}$ is obtained as
\begin{equation}
   \label{eq:S-ins-high}
  \tilde{S}_{\rm ins} = \frac{2}{K_{\rho}}
               \ln \left( \frac{\Lambda}{2 \pi T} \right) .
\end{equation}
Substitution of eq.~(\ref{eq:S-ins-high}) into eq.~(\ref{eq:pf-high}) yields
\begin{align}
  Z & \propto 2 + 8 \cos \chi \cdot \gamma^{2} \beta^{2}
         \left( \frac{2 \pi T}{\Lambda} \right)^{\frac{2}{K_{\rho}} }
                   \nonumber \\
 & \hspace{15mm} \times
         \exp \left( - \pi \left( \frac{L}{v_{\rho} K_{\rho}}
               +\frac{K_{\sigma} L}{v_{\sigma}} \right) T \right) .
\end{align}
From eq.~(\ref{eq:j-from-Z}), we finally obtain
\begin{align}
   \label{eq:jc-high}
  j(\chi)
  & = 8 eT \left(\frac{2\pi \gamma}{\Lambda} \right)^{2}
           \left( \frac{2 \pi T}{\Lambda} \right)
                    ^{2\left( \frac{1}{K_{\rho}} - 1 \right)}
                   \nonumber \\
 & \hspace{5mm} \times
           \exp \left( - \pi \left( \frac{L}{v_{\rho} K_{\rho}}
               +\frac{K_{\sigma} L}{v_{\sigma}} \right) T \right)
           \sin \chi .
\end{align}

Next, we turn to the low-temperature regime of $T \ll v_{\rm F}/L$.
In this regime, we are allowed to retain only the term with $M=0$
in eq.~(\ref{eq:pf-1}).
Thus, the partition function is given by
\begin{equation}
   \label{eq:pf-low}
 Z \propto 1 + 2 \cos \chi \cdot \gamma^{2} 
             \int_{0}^{\beta} {\rm d}\tau_{1}{\rm d}\tau_{2} \, 
               {\rm e}^{- \tilde{S}_{\rm ins}} .
\end{equation}
To evaluate $\tilde{S}_{\rm ins}$,
we approximate as
\begin{equation}
   \label{eq:approx-1}
 \frac{4 \pi}{K_{\rho} \beta} \sum_{\Lambda > \omega > 0}
    \frac{1}{\omega} \tanh \left( \frac{L \omega}{2v_{\rho}} \right)
 \approx \frac{2}{K_{\rho}} 
        \ln \left( \frac{2{\rm e}^{\gamma}}{\pi} \cdot 
        \frac{\Lambda L}{v_{\rho}}\right) ,
\end{equation}
where $\gamma \approx 0.5772$.
The second term in $\tilde{S}_{\rm ins}$ is obtained as
\begin{align}
   \label{eq:sekibun}
 & \frac{4 \pi}{K_{\rho} \beta} \sum_{\omega > 0}
     \frac{1}{\omega} {\rm cosech} \left( \frac{L \omega}{v_{\rho}} \right)
     \left(1 - \cos \omega \left( \tau_{1} - \tau_{2} \right) \right)
                   \nonumber \\
 & \hspace{20mm}
       \approx \frac{2}{K_{\rho}}
       \ln \left[ \cosh \left( \frac{\pi v_{\rho}}{2L}
                             \bigl( \tau_{1} - \tau_{2} \bigr)
                        \right) \right] .
\end{align}
Using eqs.~(\ref{eq:approx-1}) and (\ref{eq:sekibun}),
we obtain
\begin{equation}
   \label{eq:int-low}
  \int_{0}^{\beta} {\rm d}\tau_{1}{\rm d}\tau_{2} \, 
             {\rm e}^{- \tilde{S}_{\rm ins}}
      = c \frac{\beta}{\Lambda}
          \left( \frac{v_{\rho}}{\Lambda L} \right)^{\frac{2}{K_{\rho}}-1} ,
\end{equation}
where $c$ is a numerical constant of order of unity.
Substitution of eq.~(\ref{eq:int-low}) into eq.~(\ref{eq:pf-low}) yields
\begin{equation}
  Z \propto 1 + 2 \cos \chi \cdot c \frac{v_{\rho} \beta}{L}
            \left(\frac{\gamma}{\Lambda} \right)^{2}
            \left( \frac{v_{\rho}}{\Lambda L} \right)
                    ^{2\left( \frac{1}{K_{\rho}} - 1 \right)} .
\end{equation}
We finally obtain
\begin{equation}
  j(\chi) = 4 c \frac{e v_{\rho}}{L}
            \left(\frac{\gamma}{\Lambda} \right)^{2}
            \left( \frac{v_{\rho}}{\Lambda L} \right)
                    ^{2\left( \frac{1}{K_{\rho}} - 1 \right)}
            \sin \chi .
\end{equation}

This indicates that the critical current in the low-temperature regime
behaves like
$j_{\rm c} \propto (1 / L)^{2K_{\rho}^{-1}-1}$.
This result is not consistent with
Fazio \textit{et al.}'s result, $j_{\rm c}^{\rm FHO} \propto
(1 / L)^{K_{\rho}^{-1} + K_{\sigma} - 1}$.
We discuss the reason of this inconsistency in the final section.

\section{Josephson Current in the Weak-Coupling Model}

In this section we calculate the dc Josephson current
in a TL liquid of length $L$ with open boundaries,
which is weakly coupled with the left (right) superconductor at $x = 0$
($x = L$) through a tunnel junction (see Fig. 1(a)).
We show that the resulting expression of the Josephson current
is essentially equivalent to that in the interface-barrier model
studied in \S 3.

The bosonized description of a TL liquid with open boundaries
is presented by Fabrizio and Gogolin.~\cite{rf:29}
In terms of the boson operators
$a_{q}$, $a_{q}^{\dagger}$, $b_{q}$ and $b_{q}^{\dagger}$,
and the winding-number operators $N$ and $M$,
the Hamiltonian is given by
\begin{equation}
 H = \frac{\pi}{4L} \left(  \frac{v_{\rho}}{K_{\rho}}N^{2}
                          + \frac{v_{\sigma}}{K_{\sigma}}M^{2} \right)
   + \sum_{q > 0} \left( v_{\rho}q a_{q}^{\dag} a_{q}
                       + v_{\sigma}q b_{q}^{\dag} b_{q} \right) ,
\end{equation}
where $q = \pi n/L$ ($n = 1, 2, 3, \cdots$).
The winding-number operators satisfy the condition of $N + M = {\rm even}$.
The phase fields are given by
\begin{align}
   \label{eq:f-theta1}
  \Theta_{+}(x) & = \frac{\pi N}{L}x + \theta_{+}(x) , \\
   \label{eq:f-theta2}
  \Theta_{-}(x) & = \varphi_{\rho} + \theta_{-}(x) , \\
   \label{eq:f-phi1}
  \Phi_{+}(x) & = \frac{\pi M}{L}x + \phi_{+}(x) , \\
   \label{eq:f-phi2}
  \Phi_{-}(x) & = \varphi_{\sigma} + \phi_{-}(x) ,
\end{align}
where $\theta_{\pm}$ and $\phi_{\pm}$ are the nonzero-mode components,
and $\varphi_{\rho}$ and $\varphi_{\sigma}$ are the zero-mode operators
which satisfy $[N, \varphi_{\rho}] = [M, \varphi_{\sigma}] = 2 {\rm i}$
and $[N, \varphi_{\sigma}] = [M, \varphi_{\rho}] = 0$.
The nonzero-mode components are given by
\begin{align}
   \label{eq:f-theta3}
  \theta_{+}(x) & =
         \sqrt{K_{\rho}} \sum_{q>0} \sqrt{\frac{2\pi}{q L}}
              {\rm e}^{-\alpha q/2} \sin (q x)
              \left( a_{q}^{\dag} + a_{q}  \right) , \\
   \label{eq:f-theta4}
  \theta_{-}(x) & =
         \frac{\rm i}{\sqrt{K_{\rho}}} \sum_{q>0} \sqrt{\frac{2\pi}{q L}}
              {\rm e}^{-\alpha q/2} \cos (q x)
              \left( a_{q}^{\dag} - a_{q}  \right) , \\
   \label{eq:f-phi3}
  \phi_{+}(x) & =
         \sqrt{K_{\sigma}} \sum_{q>0} \sqrt{\frac{2\pi}{q L}}
              {\rm e}^{-\alpha q/2} \sin (q x)
              \left( b_{q}^{\dag} + b_{q}  \right) , \\
   \label{eq:f-phi4}
  \phi_{-}(x) & =
         \frac{\rm i}{\sqrt{K_{\rho}}} \sum_{q>0} \sqrt{\frac{2\pi}{q L}}
              {\rm e}^{-\alpha q/2} \cos (q x)
              \left( b_{q}^{\dag} - b_{q}  \right) .
\end{align}
It is clear that $\sin (q x) = 0$ when $x = 0$ and $L$,
so that $\theta_{+}(x)$ and $\phi_{+}(x)$ are pinned
at the ends of the 1D system.
That is, $\theta_{+}(0) = \theta_{+}(L) = 0$
and $\phi_{+}(0) = \phi_{+}(L) = 0$.
This means that fluctuations in both charge and spin
are strongly suppressed in the vicinity of both the ends
due to the open boundary condition.
This result plays an important role in our argument.
In terms of the phase fields,
the right-moving component $\psi_{1s}$ and the left-moving component
$\psi_{2s}$ of the electron field are expressed as~\cite{rf:29}
\begin{align}
   \label{eq:psi-1d}
   \psi_{1s}(x)
 & = \frac{1}{\sqrt{2 \pi \alpha}}
               \exp \Big( {\rm i} k_{\rm F} x + \frac{\rm i}{2} 
               \big\{ \Theta_{+}(x) + \Theta_{-}(x)
                   \nonumber \\
 & \hspace{25mm}
                        + s \Phi_{+}(x) + s \Phi_{-}(x) \big\} \Big) , \\
   \label{eq:psi-2d}
   \psi_{2s}(x)
 & = \frac{-1}{\sqrt{2 \pi \alpha}}
               \exp \Big( - {\rm i} k_{\rm F} x + \frac{\rm i}{2} 
               \big\{ -\Theta_{+}(x) + \Theta_{-}(x)
                   \nonumber \\
 & \hspace{25mm}
                         - s \Phi_{+}(x) + s \Phi_{-}(x) \big\} \Big) .
\end{align}
We neglected Majorana fermions in eqs.~(\ref{eq:psi-1d})
and (\ref{eq:psi-2d}), which do not play an essential role in our argument.

Now we introduce the Hamiltonian which describes
the superconductors and their coupling with the TL liquid.
We assume that our STLLS system is symmetric with respect to the NS contacts.
That is, the left contact at $x = 0$ is equivalent to
the right contact at $x = L$.
Thus, we present here only the part of the Hamiltonian
related with the left superconductor.
The Hamiltonian for the left superconductor is given by
\begin{align}
 H^{\rm S}
 & =  \sum_{k, s} \epsilon_{k}
                         c_{{\rm L}k, s}^{\dagger}c_{{\rm L}k, s}
                   \nonumber \\
 & \hspace{3mm}
     -\sum_{k} \left( \Delta {\rm e}^{{\rm i} \chi_{1}}
                         c_{{\rm L}k, +}^{\dagger}c_{{\rm L}-k, -}^{\dagger}
                    + \Delta {\rm e}^{-{\rm i} \chi_{1}}
                         c_{{\rm L}-k, -}c_{{\rm L}k, +}  \right) ,
\end{align}
where $\epsilon_{k}$ denotes the single electron spectrum
and $c_{{\rm L}k, s}$ ($c_{{\rm L}k, s}^{\dagger}$)
is the fermion anihilation (creation) operator.
In terms of the Bogoliubov transformation
\begin{align}
   \label{eq:bogoliubov}
 c_{{\rm L}k, +} & = u_{k} d_{{\rm L}k, +}
             + v_{k}{\rm e}^{{\rm i}\chi_{1}}d_{{\rm L}-k, -}^{\dagger} , \\
 c_{{\rm L}-k, -} & = u_{k} d_{{\rm L}-k, -}
               - v_{k}{\rm e}^{{\rm i}\chi_{1}}d_{{\rm L}k, +}^{\dagger} ,
\end{align}
with
\begin{align}
 u_{k} & = \sqrt{ \frac{1}{2}
                    \left( 1 + \frac{\epsilon_{k}}{E_{k}} \right)} , \\
 v_{k} & = \sqrt{ \frac{1}{2}
                    \left( 1 - \frac{\epsilon_{k}}{E_{k}} \right)} ,
\end{align}
the Hamiltonian is diagonalized as
\begin{equation}
   \label{eq:h-s-d}
  H^{\rm S} = \sum_{k, s} E_{k} d_{{\rm L}k, s}^{\dagger} d_{{\rm L}k, s} ,
\end{equation}
where $E_{k} = \sqrt{ \epsilon_{k}^{2} + \Delta^{2} }$.
The weak coupling between the TL liquid and the left superconductor
is described by the tunneling Hamiltonian
\begin{equation}
   \label{eq:h-t-d}
 H^{\rm T} = \frac{1}{\sqrt{V}} \sum_{k, s}
               \left( c_{{\rm L}k,s}^{\dag} \int {\rm d}x 
                        t_{{\rm L}k}(x) \psi_{s}(x)
                          + {\rm h. c.}  \right) ,
\end{equation}
where $V$ is the volume of the superconductor.
We assume that the tunneling-matrix element $t_{{\rm L}k}(x)$ has
nonzero values only in the vicinity of $x = 0$
and vanishes when $\alpha_{\rm T} < x$,
where $\alpha_{\rm T}$ represents the length scale of the order of
a few lattice spacings.
The Hamiltonian for the right superconductor is simply given by
replacing $\rm L$ with $\rm R$ in eqs.~(\ref{eq:h-s-d}) and (\ref{eq:h-t-d}).
Note that $t_{{\rm R}k}(x)$ has nonzero values only in the vicinity of
$x = L$ and vanishes when $x < L - \alpha_{\rm T}$.

In order to calculate the partition function,
it is convenient to derive the effective action $S^{\Gamma}$
describing the coupling with the TL liquid and the superconductors.
The action $S_{\Gamma}$ is obtained by integrating out the electron fields
in the superconductors,
\begin{align}
 & \int \prod_{k, s}
                 \mathcal{D}d_{{\rm L} k, s}^{\dagger}
                 \mathcal{D}d_{{\rm L} k, s}
                 \mathcal{D}d_{{\rm R} k, s}^{\dagger}
                 \mathcal{D}d_{{\rm R} k, s}
           \exp \left( - S^{\rm S} - S^{\rm T} \right)
                   \nonumber \\
 & \hspace{20mm}
     \propto \exp \left( - S^{\Gamma} \right) ,
\end{align}
where
\begin{align}
  S^{\rm S}
 & = \int_{0}^{\beta} {\rm d} \tau \sum_{k, s}
           \Big(
                d_{{\rm L}k, s}^{\dagger} \partial_{\tau} d_{{\rm L}k, s}
                      + E_{k} d_{{\rm L}k, s}^{\dagger} d_{{\rm L}k, s}
                   \nonumber \\
 & \hspace{20mm}
              + d_{{\rm R}k, s}^{\dagger} \partial_{\tau} d_{{\rm R}k, s}
                      + E_{k} d_{{\rm R}k, s}^{\dagger} d_{{\rm R}k, s} 
           \Big) , \\
  S^{\rm T}
 & = \int_{0}^{\beta} {\rm d} \tau \frac{1}{\sqrt{V}} \sum_{k, s}
           \Big(  c_{{\rm L}k, s}^{\dag}
                       \int {\rm d}x t_{{\rm L}k}(x) \psi_{s}(x)
                   \nonumber \\
 & \hspace{20mm}
                 + c_{{\rm R}k, s}^{\dag}
                       \int {\rm d}x t_{{\rm R}k}(x) \psi_{s}(x)
                          + {\rm h. c.}  \Big) .
\end{align}
The derivation of $S^{\Gamma}$ is outlined in Appendix C.
The result is
\begin{align}
  \label{eq:s-Gamma}
 S^{\Gamma} & = \int_{0}^{\beta} {\rm d}\tau_{1} {\rm d}\tau_{2}
                   Q(\tau_{1} - \tau_{2})
        \nonumber \\
    & \hspace{5mm} \times
          \Big(  {\rm e}^{-{\rm i}\chi_{1}}
                    \big\{  \psi_{1-}(0, \tau_{1}) \psi_{2+}(0, \tau_{2})
                   \nonumber \\
    & \hspace{17mm}
                          + \psi_{2-}(0, \tau_{1}) \psi_{1+}(0, \tau_{2})
                    \big\} + {\rm h. c.}
                 \nonumber \\
    & \hspace{10mm}
             + {\rm e}^{-{\rm i}\chi_{2}}
                    \big\{  \psi_{1-}(L, \tau_{1}) \psi_{2+}(L, \tau_{2})
                   \nonumber \\
    & \hspace{17mm}
                          + \psi_{2-}(L, \tau_{1}) \psi_{1+}(L, \tau_{2})
                    \big\}) + {\rm h. c.}
          \Big) ,
\end{align}
where $Q(\tau)$ is defined in Appendix C.
We substitute eqs.~(\ref{eq:psi-1d}) and (\ref{eq:psi-2d}) into
eq.~(\ref{eq:s-Gamma}).
Noting that $\theta_{+}(0) = \theta_{+}(L) = 0$
and $\phi_{+}(0) = \phi_{+}(L) = 0$, we obtain
\begin{align}
  \label{eq:s-Gamma-w}
 S^{\Gamma} & = - \frac{1}{\pi \alpha}
                  \int_{0}^{\beta} {\rm d}\tau_{1} {\rm d}\tau_{2}
                           Q(\tau_{1} - \tau_{2})   \nonumber \\
    & \times
        \bigg[  {\rm e}^{-{\rm i}\chi_{1}}
                   \exp \frac{\rm i}{2} \Big( \Theta_{-}(0, \tau_{1})
                                              + \Theta_{-}(0, \tau_{2})
                   \nonumber \\
    & \hspace{17mm}
                                              - \Phi_{-}(0, \tau_{1})
                                              + \Phi_{-}(0, \tau_{2})
                                        \Big) + {\rm c. c.}
             \nonumber \\
    & \hspace{0mm}
             + (-1)^{M} {\rm e}^{-{\rm i}\chi_{2}}
                   \exp \frac{\rm i}{2} \Big(  \Theta_{-}(L, \tau_{1})
                                               + \Theta_{-}(L, \tau_{2})
                   \nonumber \\
    & \hspace{17mm}
                                               - \Phi_{-}(L, \tau_{1})
                                               + \Phi_{-}(L, \tau_{2})
                                        \Big) + {\rm c. c.}
        \bigg] .
\end{align}
The kernel $Q(\tau)$ vanishes when $|\tau| \gg \Delta^{-1}$
as shown in Appendix C,
while the characteristic time scale of $\Theta_{+}$ and $\Phi_{-}$
is much longer than $\Delta^{-1}$.
Thus, we can approximate in eq.~(\ref{eq:s-Gamma-w}) that
\begin{equation}
   Q(\tau_{i} - \tau_{j}) = \Gamma \delta(\tau_{i} - \tau_{j}) ,
\end{equation}
where $\Gamma = \int {\rm d} \tau Q(\tau)$.
Thus, $S^{\Gamma}$ is simplified to
\begin{align}
  \label{eq:s-Gamma-x}
 S^{\Gamma}
 & = - \frac{\Gamma}{\pi \alpha}
                  \int_{0}^{\beta} {\rm d}\tau
       \bigg[ \Big( {\rm e}^{-{\rm i}\chi_{1} + {\rm i} \Theta_{-}(0, \tau)}
                             + {\rm c. c.} \Big)
                   \nonumber \\
 & \hspace{15mm}
   + (-1)^{M} \Big( {\rm e}^{-{\rm i}\chi_{2} + {\rm i} \Theta_{-}(L, \tau)}
                             + {\rm c. c.} \Big)
      \bigg] . 
\end{align}
The sign of the second term depends on $M$,
so we express $S^{\Gamma}$ with even $M$ and that of odd $M$
as $S_{+}^{\Gamma}$ and $S_{-}^{\Gamma}$, respectively.

Note that $S_{\pm}^{\Gamma}$ does not contain $\varphi_{\sigma}$,
so that $M$ is a conserved quantity.
By contrast, $N$ is no longer conserved in the presence of $S_{\pm}^{\Gamma}$.
The partition function is expressed as
\begin{align}
   \label{eq:pf-1d}
  Z & = \sum_{M = - \infty \atop (M = {\rm even})}^{\infty}
         \exp \left(-\beta \frac{\pi v_{\sigma} M^{2}}{4K_{\sigma}L}
              \right) Z_{+}
                   \nonumber \\
    & \hspace{15mm}
    + \sum_{M = - \infty \atop (M = {\rm odd})}^{\infty}
         \exp \left(-\beta \frac{\pi v_{\sigma} M^{2}}{4K_{\sigma}L}
              \right) Z_{-} ,
\end{align}
with
\begin{align}
  Z_{\pm}
 & = \sum_{m = -\infty}^{\infty}
            \int_{\Delta \varphi_{\rho} = 2\pi m}
                \mathcal{D} \varphi_{\rho}
            \int \prod_{q>0}
                 \mathcal{D}a_{q}^{\dagger} \mathcal{D}a_{q}
                   \nonumber \\
 & \hspace{5mm} \times
            \exp \left( {\rm i}\int_{0}^{\beta} {\rm d} \tau 
                         \frac{\zeta_{\pm}}{2 \pi}\partial_{\tau}
                                  \varphi_{\rho}
                       - S^{\rm Z} - S^{\rm NZ} - S_{\pm}^{\Gamma}
                 \right) ,
\end{align}
where $\zeta_{+} = 0$ and $\zeta_{-} = \pi$, and
\begin{align}
  S^{\rm Z} & = \int_{0}^{\beta} {\rm d} \tau
                          \frac{L}{4 \pi v_{\rho} K_{\rho}}
                          (\partial_{\tau} \varphi_{\rho})^{2} , \\
  S^{\rm NZ} & = \int_{0}^{\beta} {\rm d} \tau \sum_{q>0}
                     \left( a_{q}^{\dagger} \partial_{\tau} a_{q}
                              + v_{\rho} q a_{q}^{\dagger} a_{q} \right) .
\end{align}
We introduced $\zeta_{\pm}$ to maintain the constraint
$N + M = \mbox{even}$.~\cite{rf:33}
Since it is assumed that the coupling between the TL liquid and
the superconductors is weak,
we calculate the partition function by treating $S_{\pm}^{\Gamma}$
as a perturbation.
Within the lowest-order approximation with respect to $S_{\pm}^{\Gamma}$,
we find that $Z_{\pm} \propto 1 + Z_{1\pm}/Z_{0\pm}$, where
\begin{align}
 Z_{0\pm}
  & = \sum_{m = -\infty}^{\infty}
              \int_{\Delta \varphi_{\rho} = 2\pi m}
                  \mathcal{D} \varphi_{\rho}
              \int \prod_{q>0}
                  \mathcal{D}a_{q}^{\dagger} \mathcal{D}a_{q}
                   \nonumber \\
 & \hspace{25mm} \times
              \exp \left( {\rm i}m \zeta_{\pm}
                       - S^{\rm Z} - S^{\rm NZ} \right) , \\
 Z_{1\pm}
  & = \sum_{m = -\infty}^{\infty}
              \int_{\Delta \varphi_{\rho} = 2\pi m}
                 \mathcal{D} \varphi_{\rho}
              \int \prod_{q>0}
                 \mathcal{D}a_{q}^{\dagger} \mathcal{D}a_{q}
              \frac{1}{2} \left( S_{\pm}^{\Gamma} \right)^{2}
                   \nonumber \\
 & \hspace{25mm} \times
              \exp \left( {\rm i}m \zeta_{\pm}
                       - S^{\rm Z} - S^{\rm NZ} \right) .
\end{align}
Using eq.~(\ref{eq:s-Gamma-x}), we find that
\begin{align}
       \label{eq:Z1-pm-d}
 Z_{1\pm}
  & = \pm \sum_{m = -\infty}^{\infty} {\rm e}^{{\rm i}m \zeta_{\pm}}
           \int_{\Delta \varphi_{\rho} = 2\pi m} \mathcal{D} \varphi_{\rho}
              \int \prod_{q>0} \mathcal{D}a_{q}^{\dagger} \mathcal{D}a_{q}
                   \nonumber \\
 & \hspace{5mm} \times
        \exp \left( - S^{\rm Z} - S^{\rm NZ} \right)
          \nonumber \\
 & \hspace{5mm} \times
        \left( \frac{\Gamma}{\pi \alpha} \right)^{2}
        \int_{0}^{\beta} {\rm d} \tau_{1} {\rm d} \tau_{2}
          \nonumber \\
 & \hspace{5mm} \times
        \Biggl(  {\rm e}^{{\rm i}\chi}
                   \exp {\rm i} \Bigl(  \Theta_{-}(0, \tau_{1})
                                      - \Theta_{-}(L, \tau_{2})
                                \Bigr)
                                          + {\rm c. c.} \Biggr) ,
\end{align}
where $\chi = \chi_{2} - \chi_{1}$
and we neglected terms which are independent of $\chi$.
To proceed, we decompose $\varphi_{\rho}$ as
$\varphi_{\rho}(\tau) = \varphi_{\rho}^{0} + 2\pi m \tau / \beta
+ \tilde{\varphi}_{\rho}(\tau)$,
where $\tilde{\varphi}_{\rho}(\beta) - \tilde{\varphi}_{\rho}(0) = 0$.
Using this decomposition, we obtain
\begin{align}
 Z_{0\pm}
  & = \sum_{m = -\infty}^{\infty}
           {\rm e}^{- \frac{\pi K_{\rho} L}{v_{\rho}\beta }m^{2}}
             \int \mathcal{D} \tilde{\varphi}_{\rho}
             \int \prod_{q>0} \mathcal{D}a_{q}^{\dagger} \mathcal{D}a_{q}
          \nonumber \\
 & \hspace{5mm} \times
          \exp \left( - S^{\rm Z} - S^{\rm NZ} \right) , \\
 Z_{1\pm}
  & = \pm \left( \frac{\Gamma}{\pi \alpha} \right)^{2}
           \int_{0}^{\beta} {\rm d} \tau_{1} {\rm d} \tau_{2}
        \sum_{m = -\infty}^{\infty}
          \nonumber \\
 & \hspace{5mm} \times
           {\rm e}^{- \frac{\pi K_{\rho} L}{v_{\rho}\beta }m^{2}}
           {\rm e}^{{\rm i}m \zeta_{\pm}
                     + {\rm i} \frac{2\pi m}{\beta}(\tau_{1} - \tau_{2}) }
          \nonumber \\
  & \hspace{5mm} \times
           \int \mathcal{D} \tilde{\varphi}_{\rho}
              \int \prod_{q>0} \mathcal{D}a_{q}^{\dagger} \mathcal{D}a_{q}
        \exp \left( - S^{\rm Z} - S^{\rm NZ} \right)
          \nonumber \\
  & \hspace{5mm} \times
        \Bigg\{ {\rm e}^{{\rm i}\chi}
                \exp {\rm i} \Bigl(  \tilde{\varphi}_{\rho}(\tau_{1})
                                   - \tilde{\varphi}_{\rho}(\tau_{2})
                                   + \theta_{-}(0, \tau_{1})
          \nonumber \\
 & \hspace{35mm}
                                    - \theta_{-}(L, \tau_{2})
                             \Bigr)
                                    + {\rm c. c.} \Bigg\} .
\end{align}
To evaluate $Z_{1\pm}$, it is convenient to introduce
the correlation function defined by
\begin{align}
  \Omega \left( \tau_{1} - \tau_{2} \right)
 & = \Bigm\langle \exp {\rm i}
     \Big(  \tilde{\varphi}_{\rho}(\tau_{1})
           - \tilde{\varphi}_{\rho}(\tau_{2})
          \nonumber \\
  & \hspace{15mm}
           + \theta_{-}(0, \tau_{1})
           - \theta_{-}(L, \tau_{2}) \Big) \Bigm\rangle ,
\end{align}
where
\begin{equation}
  \langle \cdots \rangle
 = \frac{ \int \mathcal{D} \tilde{\varphi}_{\rho}
              \int \prod_{q>0} \mathcal{D}a_{q}^{\dagger} \mathcal{D}a_{q}
            \cdots \exp \left( - S^{\rm Z} - S^{\rm NZ} \right) }
        { \int \mathcal{D} \tilde{\varphi}_{\rho}
              \int \prod_{q>0} \mathcal{D}a_{q}^{\dagger} \mathcal{D}a_{q}
            \exp \left( - S^{\rm Z} - S^{\rm NZ} \right) } .
\end{equation}
The correlation function is obtained as
\begin{align}
  \Omega \left( \tau \right)
  & = \exp \bigg(  - \frac{\pi L}{K_{\rho}v_{\rho} \beta}
               - \frac{4 \pi}{K_{\rho} \beta} \sum_{\Lambda > \omega > 0}
                     \frac{1}{\omega}
                     \tanh \left( \frac{L \omega}{2v_{\rho}} \right)
              \nonumber \\
  & \hspace{1mm} 
                 - \frac{4 \pi}{K_{\rho} \beta} \sum_{\omega > 0}
                     \frac{1}{\omega}
                     {\rm cosech} \left( \frac{L \omega}{v_{\rho}} \right)
                      \left( 1 - \cos \omega \tau \right)
       \bigg) ,
\end{align}
where the high-frequency cutoff $\Lambda$ is introduced
for the second term in the exponent.
By using $\Omega$, we can express $Z_{1\pm} / Z_{0\pm}$ as
\begin{align}
  \frac{Z_{1\pm}}{Z_{0\pm}}
 & = \pm 2 \gamma^{2} \cos \chi
           \int_{0}^{\beta} {\rm d} \tau_{1} {\rm d} \tau_{2}
           \nonumber \\
 & \hspace{5mm} \times
        \frac{ \sum_{m}
                {\rm e}^{- \frac{\pi K_{\rho} L}{v_{\rho}\beta }m^{2}}
                {\rm e}^{{\rm i}m \zeta_{\pm}
                     + {\rm i} \frac{2\pi m}{\beta}(\tau_{1} - \tau_{2})} }
             { \sum_{m}
                {\rm e}^{- \frac{\pi K_{\rho} L}{v_{\rho}\beta }m^{2}} }
          \Omega \left( \tau_{1} - \tau_{2} \right) ,
\end{align}
where we set $\gamma = \Gamma /(\pi \alpha)$.

We first calculate the Josephson current in the high-temperature regime
of $v_{\rho} / L \ll T \ll \Delta$.
In this regime, we are allowed to retain only the term with $m = 0$,
and obtain
\begin{equation}
 \frac{Z_{1\pm}}{Z_{0\pm}} =
       \pm 2 \gamma^{2} \cos \chi
           \int_{0}^{\beta} {\rm d} \tau_{1} {\rm d} \tau_{2}
           \Omega \left( \tau_{1} - \tau_{2} \right) ,
\end{equation}
where
\begin{equation}
   \Omega \left( \tau \right) \approx
      \left( \frac{2\pi T}{\Lambda} \right)^{\frac{2}{K_{\rho}}}
        \exp \left( - \frac{\pi L}{K_{\rho}v_{\rho}} T \right) .
\end{equation}
The partition function is
\begin{align}
   \label{eq:ZZ-high}
 Z & \propto
      \left( 1 + 2 {\rm e}^{- \frac{\pi K_{\sigma}L}{v_{\sigma}}T} \right)
          \left( 1 + \frac{Z_{1+}}{Z_{0+}} \right)
          \nonumber \\
  & \hspace{15mm}
    + \left( 1 - 2 {\rm e}^{- \frac{\pi K_{\sigma}L}{v_{\sigma}}T} \right)
          \left( 1 - \frac{Z_{1+}}{Z_{0+}} \right)
           \nonumber \\
   & = 2 + 8 \cos \chi \cdot \gamma^{2} \beta^{2}
          \left( \frac{2\pi T}{\Lambda} \right)^{\frac{2}{K_{\rho}}} 
          \nonumber \\
  & \hspace{7mm} \times
          \exp \left( - \pi \left(  \frac{L}{K_{\rho}v_{\rho}}
                                  + \frac{K_{\sigma}L}{v_{\sigma}}
                            \right) T \right) .
\end{align}
We have used eqs.~(\ref{eq:sum-M1}) and (\ref{eq:sum-M2})
in deriving eq.~(\ref{eq:ZZ-high}).
The Josephson current is obtained as
\begin{align}
 j (\chi)
    & = -2 e T \frac{\partial \ln Z}{\partial \chi}
                   \nonumber \\
    & = 8 eT \left( \frac{2\pi \gamma}{\Lambda} \right)^{2}
               \left( \frac{2\pi T}{\Lambda} \right)
                        ^{2 \left( \frac{1}{K_{\rho}}-1 \right)}
          \nonumber \\
    & \hspace{5mm} \times
               \exp \left(  - \pi \left(  \frac{L}{K_{\rho}v_{\rho}}
                                        + \frac{K_{\sigma}L}{v_{\sigma}}
                                  \right) T \right)
               \sin \chi .
\end{align}
This result is equivalent to that obtained in \S 3
in the high-temperature regime.

Next we consider the low-temperature regime of $T \ll v_{\rho}/L$,
where the terms with $M = 0$ dominantly contribute to $Z$.
We then find that $Z \propto 1 + Z_{1+}/Z_{0+}$.
Using Poisson's summation theorem,
we rewrite the summation over $m$ in $Z_{1+}$,
\begin{align}
  & \sum_{m = - \infty}^{\infty}
      {\rm e}^{- \frac{\pi K_{\rho} L}{v_{\rho}\beta }m^{2}}
      {\rm e}^{{\rm i} \frac{2\pi m}{\beta}(\tau_{1} - \tau_{2})}
          \nonumber \\
  & \hspace{5mm}
    \propto
    \sum_{q = - \infty}^{\infty}
      \exp \left( - \frac{v_{\rho} \beta}{4 \pi K_{\rho}L}
                 \left( 2\pi q + \frac{2\pi}{\beta}
                           \left( \tau_{1} - \tau_{2} \right) \right)^{2}
           \right) .
\end{align}
This equation indicates that we are allowed to retain only
the term with $q = 0$.
Then, we obtain
\begin{align}
 \frac{Z_{1+}}{Z_{0+}}
  & = 2 \cos \chi \cdot \gamma^{2} \beta^{2}
         \int_{0}^{\beta} {\rm d} \tau_{1} {\rm d} \tau_{2}
         {\rm e}^{- \frac{\pi v_{\rho}}{K_{\rho} L \beta}
                 \left( \tau_{1} - \tau_{2} \right)^{2} }
          \nonumber \\
  & \hspace{35mm} \times
         \Omega \left( \tau_{1} - \tau_{2} \right) ,
\end{align}
with
\begin{equation}
 \Omega \left( \tau \right)
   \approx c' \left( \frac{v_{\rho}}{\Lambda L} \right)^{\frac{2}{K_{\rho}}}
                    \cosh \left( \frac{\pi v_{\rho}}{2L} \tau \right)
                                  ^{-\frac{2}{K_{\rho}}} ,
\end{equation}
where $c'$ is a numerical constant of order of unity.
The partition function is obtained as
\begin{equation}
 Z \propto 1 + 2 \cos \chi \cdot c \gamma^{2} \frac{\beta}{\Lambda}
       \left( \frac{v_{\rho}}{\Lambda L} \right)^{\frac{2}{K_{\rho}} - 1} ,
\end{equation}
where $c$ is a numerical constant of order of unity.
We finally obtain
\begin{align}
 j (\chi)
    & = - 2 e T \frac{\partial \ln Z}{\partial \chi}
                   \nonumber \\
    & =   4 c \frac{e v_{\rho}}{L}
            \left(\frac{\gamma}{\Lambda} \right)^{2}
            \left( \frac{v_{\rho}}{\Lambda L} \right)
                    ^{2 \left( \frac{1}{K_{\rho}} - 1 \right)}
            \sin \chi .
\end{align}
The result is also equivalent to
that obtained in \S 3 in the low-temperature regime.

\section{Discussion}

We studied the dc Josephson effect in a TL liquid
on the basis of the two different models.
We derived a bosonized description of
a superconductor--TL-liquid--superconductor (STLLS) system
with perfect NS interfaces, and calculate the Josephson current
by introducing a strong barrier at each NS interface.
We next examined the Josephson current through a TL liquid
with open boundaries,
where the TL liquid is weakly coupled at its left (right) end
with the left (right) superconductor through a tunnel junction.
It is shown that the two models provide us the same expression of
the Josephson current
in both the high-temperature regime of $v_{\rm F}/L \ll T \ll \Delta$
and the low-temperature regime of $T \ll v_{\rm F}/L$.
We found that the Josephson current in the high-temperature regime
is given by
\begin{align}
  j(\chi)
 & = 8 eT \left(\frac{2\pi \gamma}{\Lambda} \right)^{2}
          \left( \frac{2 \pi T}{\Lambda} \right)
                     ^{2\left( \frac{1}{K_{\rho}} - 1 \right)}
          \nonumber \\
  & \hspace{10mm} \times
          \exp \left( - \pi \left( \frac{L}{v_{\rho} K_{\rho}}
               +\frac{K_{\sigma} L}{v_{\sigma}} \right) T \right)
          \sin \chi ,
\end{align}
while
\begin{equation}
  j(\chi) = 4 c \frac{e v_{\rho}}{L}
            \left(\frac{\gamma}{\Lambda} \right)^{2}
            \left( \frac{v_{\rho}}{\Lambda L} \right)
                    ^{2\left( \frac{1}{K_{\rho}} - 1 \right)}
            \sin \chi ,
\end{equation}
in the low-temperature regime.

We treated the strong-barrier limit of the interface-barrier model in \S 3.
Let us briefly consider the opposite weak-barrier case.
For simplicity, both the barriers are assumed to have the same strength.
We employ the notations presented in \S 3.
The action describing the weak barrier potential is
\begin{equation}
  S^{\rm B} = - \frac{4 V_{0}}{\pi \alpha} \int {\rm d} \tau
              \cos (\varphi_{\rho} + \bar{\theta})
              \cos \frac{\tilde{\theta}}{2} ,
\end{equation}
with $V_{0}/(\alpha \Lambda) \ll 1$.
We have used the fact that the spin degrees of freedom are frozen
in the vicinity of the NS interfaces.
The potential becomes large after renormalization.
We apply a perturbative renormalization-group argument to obtain
a scaling equation for the barrier potential.~\cite{rf:6, rf:7, rf:8,
rf:9, rf:10}
If the band width $\Lambda$ is reduced to $\mu$,
the potential becomes
\begin{equation}
 V_{0}(\mu) = V_{0}(\Lambda)
        \left( \frac{\mu}{\Lambda} \right)^{K_{\rho} - 1} ,
\end{equation}
as long as $v_{\rho}/L \ll \mu < \Lambda$.
This means that the potential is renormalized to be large
in the repulsive interaction cases of $K_{\rho} < 1$.
Thus, the system flows to the strong-barrier limit,
where our analysis in \S 3 is justified.
Since the coupling between a TL liquid and superconductors is
reduced by the strong barriers in this limit,
we expect that the weak-coupling model also describes
correct low-energy physics.
Our results given in \S 3 and \S 4 support this reasoning.

To elucidate characteristic features of the interface-barrier model,
it is instructive to compare our problem with
the double-barrier problem in a TL liquid.
Kane and Fisher,~\cite{rf:7}
and subsequently Furusaki and Nagaosa,~\cite{rf:9}
studied transport properties of a TL liquid
in the presence of a double-barrier structure,
which consists of two $\delta$-function barriers at $x = 0$ and $L$.
This system is similar to our Josephson system
except for the absence of superconductors.
The Hamiltonian of the double-barrier problem is given by
$H = H^{\rm 1D} + H^{\rm B}$, where
\begin{align}
  H^{\rm 1D}
 & = \int_{- \infty}^{\infty} {\rm d} x
         \left( \frac{v_{\rho}}{4\pi K_{\rho}}
                    \left( \frac{\partial \theta_{+}}{\partial x} \right)^{2}
              + \frac{K_{\rho}v_{\rho}}{4\pi}
                    \left( \frac{\partial \theta_{-}}{\partial x} \right)^{2}
         \right) \nonumber \\
 & +  \int_{- \infty}^{\infty} {\rm d} x
         \left( \frac{v_{\sigma}}{4\pi K_{\sigma}}
                    \left( \frac{\partial \phi_{+}}{\partial x} \right)^{2}
              + \frac{K_{\sigma}v_{\sigma}}{4\pi}
                    \left( \frac{\partial \phi_{-}}{\partial x} \right)^{2}
         \right) , \\
    \label{eq:disc-H^B}
  H^{\rm B}
 & = - \frac{4 V_{0}}{\pi \alpha} \Bigg[
         \cos \bar{\theta} \cos \frac{\tilde{\theta}}{2}
         \cos \bar{\phi} \cos \frac{\tilde{\phi}}{2}
          \nonumber \\
  & \hspace{25mm}
     + \sin \bar{\theta} \sin \frac{\tilde{\theta}}{2}
         \sin \bar{\phi} \sin \frac{\tilde{\phi}}{2}
                                           \Bigg]  .
\end{align}
We assumed that $k_{\rm F} L = 0$ (mod $\pi $).
The variables $\bar{\theta}$, $\tilde{\theta}$, $\bar{\phi}$ and
$\tilde{\phi}$ in eq.~(\ref{eq:disc-H^B}) are defined as
$\bar{\theta} = (\theta_{+}(L) + \theta_{+}(0))/2$,
$\tilde{\theta} = \theta_{+}(L) - \theta_{+}(0)$,
$\bar{\phi} = (\phi_{+}(L) + \phi_{+}(0))/2$
and $\tilde{\phi} = \phi_{+}(L) - \phi_{+}(0)$, respectively.
Note that $\theta_{-}$ and $\phi_{-}$ are not contained in $H^{\rm B}$.
Thus, the partition function is
\begin{equation}
 Z = \int \mathcal{D} \theta_{+}
     \int \mathcal{D} \phi_{+}
     \exp \left( - S^{\rm 1D} - S^{\rm B} \right) ,
\end{equation}
where
\begin{align}
  S^{\rm 1D}
 & = \int_{0}^{\beta} {\rm d} \tau \int_{-\infty}^{\infty} {\rm d}x
         \Bigg( \frac{v_{\rho}}{4\pi K_{\rho}}
                 \left( \frac{\partial \theta_{+}}{\partial x} \right)^{2}
           \nonumber \\
 & \hspace{30mm}
              + \frac{1}{4\pi K_{\rho} v_{\rho}}
                 \left( \frac{\partial \theta_{+}}{\partial \tau} \right)^{2}
         \Bigg) \nonumber \\
 & +  \int_{0}^{\beta} {\rm d} \tau \int_{- \infty}^{\infty} {\rm d} x
         \Bigg( \frac{v_{\sigma}}{4\pi K_{\sigma}}
                 \left( \frac{\partial \phi_{+}}{\partial x} \right)^{2}
          \nonumber \\
 & \hspace{30mm}
              + \frac{1}{4\pi K_{\sigma} v_{\sigma}}
                 \left( \frac{\partial \phi_{+}}{\partial \tau} \right)^{2}
         \Bigg) , \\
  S^{\rm B}
 & = \int_{0}^{\beta} {\rm d} \tau H^{\rm B}(\tau) .
\end{align}
We can simplify the partition function by using the effective action
for $\bar{\theta}$, $\tilde{\theta}$, $\bar{\phi}$ and $\tilde{\phi}$.
The result is~\cite{rf:9}
\begin{equation}
 Z = \int \mathcal{D} \bar{\theta} \mathcal{D} \tilde{\theta}
     \int \mathcal{D} \bar{\phi} \mathcal{D} \tilde{\phi}
     \exp \left( - S_{\rm eff}^{\rm 1D} - S^{\rm B} \right) ,
\end{equation}
where
\begin{align}
  \label{eq:S-eff-discussion}
   S_{\rm eff}^{\rm 1D}
 & =        \frac{1}{2 \pi K_{\rho} \beta} \sum_{\omega}
            \frac{2 |\omega|}{1 + \exp(-L |\omega| /v_{\rho})}
            \bar{\theta}(\omega) \bar{\theta}(-\omega)
          \nonumber \\
 & \hspace{5mm}
          + \frac{1}{8 \pi K_{\rho} \beta} \sum_{\omega}
            \frac{2 |\omega|}{1 - \exp(-L |\omega| /v_{\rho})}
            \tilde{\theta}(\omega) \tilde{\theta}(-\omega)
          \nonumber \\
 & \hspace{5mm}
          + \frac{1}{2 \pi K_{\sigma} \beta} \sum_{\omega}
            \frac{2 |\omega|}{1 + \exp(-L |\omega| /v_{\sigma})}
            \bar{\phi}(\omega) \bar{\phi}(-\omega)
          \nonumber \\
 & \hspace{5mm}
          + \frac{1}{8 \pi K_{\sigma} \beta} \sum_{\omega}
            \frac{2 |\omega|}{1 - \exp(-L |\omega| /v_{\sigma})}
            \tilde{\phi}(\omega) \tilde{\phi}(-\omega) .
         \nonumber \\
\end{align}
We can clarify the influence of superconductors on a TL liquid
by comparing the effective action $S_{\rm eff}^{\rm NZ}$ given in
eq.~(\ref{eq:S-eff-NZ})
with that of the double-barrier problem $S_{\rm eff}^{\rm 1D}$
given in eq.~(\ref{eq:S-eff-discussion}).
As noted in \S 3,
we see from eqs.~(\ref{eq:bar-J-sigma}) and (\ref{eq:tilde-J-sigma})
that $\bar{J}_{\sigma}^{-1}(\omega)$ and
$\tilde{J}_{\sigma}^{-1}(\omega)$ are of the order of $\Delta$
in the limit of $\omega \to 0$.
This indicates that $\bar{\phi}$ and $\tilde{\phi}$ have a mass gap
of the order of $\Delta$ in our problem,
while such large gap does not appear in the double-barrier problem.
We can understand the feature by noting that
only a spin-singlet pair of two electrons can transfer the NS interfaces
due to Andreev reflection.
The conservation of the total spin through the Andreev reflection process
leads to suppression of the spin fluctuations near the interfaces.
The mass gap of the order of $\Delta$ reflects it.
The presence of superconductors also affects the charge degrees of freedom.
We also see that $\bar{J}_{\rho}^{-1}(\omega)$ and
$\tilde{J}_{\rho}^{-1}(\omega)$ are of the order of $v_{\rho}/L$
in the limit of $\omega \to 0$.
Thus, $\bar{\theta}$ and $\tilde{\theta}$ have a small mass gap
of the order of $v_{\rm F}/L$.
This gap is simply induced by a finite-size effect.
Let us focus our attention on the regime of
$v_{\rho}/L \ll |\omega|$,
where the finite-size effect plays no role.
The part of $S_{\rm eff}^{\rm NZ}$ describing the dynamics of
$\bar{\theta}(\omega)$ with $v_{\rho}/L \ll |\omega|$ is given by
\begin{equation}
  \label{eq:former}
 S_{\rm eff}^{\rm NZ} [\bar{\theta}]
   = \frac{1}{2 \pi K_{\rho} \beta }
     \sum_{v_{\rho}/L \ll |\omega| } |\omega|
     \bar{\theta}(\omega) \bar{\theta}(-\omega) .
\end{equation}
Similarly, the part of $S_{\rm eff}^{\rm 1D}$ is
\begin{equation}
  \label{eq:latter}
 S_{\rm eff}^{\rm 1D} [\bar{\theta}]
   = \frac{1}{2 \pi K_{\rho} \beta}
     \sum_{v_{\rho}/L \ll |\omega| } 2 |\omega|
     \bar{\theta}(\omega) \bar{\theta}(- \omega) .
\end{equation}
This indicates that the fluctuations in $\bar{\theta}$ are enhanced
by a factor of two in our Josephson system compared with
the double-barrier system.
The fluctuations of $\tilde{\theta}$ are also enhanced by a factor of two.
This feature is also attributed to Andreev reflection.
The reason is that the unit of charge transfer is not $e$ but $2e$
in the Andreev reflection process.
This doubling of the unit charge enhances the charge fluctuations
in the vicinity of the NS interfaces.

We obtained that the critical current is 
$j_{\rm c} \propto (1 / L)^{2 K_{\rho}^{-1} - 1}$
when $T \ll v_{\rm F}/L$.
As noted in \S 3, our result is not consistent with
Fazio \textit{et al.}'s result,~\cite{rf:19, rf:20}
$j_{\rm c}^{\rm FHO} \propto
(1 / L)^{K_{\rho}^{-1} + K_{\sigma} - 1}$.
The reason of this inconsistency is now clear.
Fazio \textit{et al.} treated a TL liquid of infinite length,
which is weakly coupled with the left (right) superconductor
at $x = 0$ ($x = L$),
under the assumption that the potential induced by the coupling
with the superconductors can be completely neglected.
Thus, within their assumption, the dynamics of the phase fields in the 1D
system is completely unaffected by the coupling.
However, even a very weak potential becomes large after renormalization
due to the repulsive electron-electorn interactions
in a TL liquid,~\cite{rf:6}
so that the TL liquid is effectively disconnected at $x = 0$ and $x = L$.
In such a situation,
$\theta_{+}$ and $\phi_{+}$ are pinned at the disconnected points,
while fluctuations in $\theta_{-}$ and $\phi_{-}$ are enhanced
around there.
From the above argument,
we see that Fazio \textit{et al.}'s result may apply to only
restricted situations,
where the induced potential is completely negligible.
If the potential becomes very large after the renormalization,
Fazio \textit{et al.}'s model must be replaced with our weak-coupling model,
as suggested by Fabrizio and Gogolin.~\cite{rf:29}
This means that our result is much general compared with
Fazio \textit{et al.}'s result.

Details of the difference between
$j_{\rm c} \propto (1/ L)^{2K_{\rho}^{-1} - 1}$
and $j_{\rm c}^{\rm FHO} \propto (1/ L)^{K_{\rho}^{-1}+K_{\sigma} - 1}$
are explained as follows.
The dynamics of the phase fields is assumed
to be unaffected by the coupling with the superconductors in refs. 19 and 20,
while, in our weak-coupling model with the open-boundary condition,
we explicitly incorporate the strong pinning of $\phi_{+}$
and the enhanced low-energy fluctuations of $\theta_{-}$
in the vicinity of the NS interfaces.
The correlation exponent $K_{\sigma}$ does not appear in the expression of
$j_{\rm c}$ due to the pinning of $\phi_{+}$,
while $j_{\rm c}^{\rm FHO}$ contains $K_{\sigma}$.
The prefactor of $K_{\rho}^{-1}$ in the exponent of $j_{\rm c}$
is doubled compared with $j_{\rm c}^{\rm FHO}$
due to the enhancement in the fluctuations of $\theta_{-}$.

\appendix
\section{Bosonization of TL liquid sandwiched between superconductors}

In this appendix, we adapt the bosonization method to
a TL liquid sandwiched between superconductors.
We first bosonize the 1D electron system in the noninteracting limit,
and then incorporate the influence of electron-electon interactions.
Our bosonization precedure is based on the argument by Haldane.~\cite{rf:17}

In terms of the eigenfunctions of the BdG equation,
presented in eqs.~(\ref{eq:eigen-f1}) and (\ref{eq:eigen-f2}),
the electron-field operators defined in eq.~(\ref{eq:fop-de}) are
expressed as~\cite{rf:30}
\begin{align}
   \label{eq:psi-1+}
 \psi_{1+}(x) & = \sum_{q > 0}
        \left( u_{q}(x) c_{q, +}
             -  g_{q}^{*}(x) c_{q, -}^{\dagger} \right) , \\
   \label{eq:psi-2+}
 \psi_{2+}(x) & = \sum_{q > 0}
        \left( f_{q}(x) d_{q, +}
             - v_{q}^{*}(x) d_{q, -}^{\dagger} \right) , \\
   \label{eq:psi-1-}
 \psi_{1-}(x) & = \sum_{q > 0}
        \left( u_{q}(x) d_{q, -}
             + g_{q}^{*}(x) d_{q, +}^{\dagger} \right) ,\\
   \label{eq:psi-2-}
 \psi_{2-}(x) & = \sum_{q > 0}
        \left( f_{q}(x) c_{q, -}
             - v_{q}^{*}(x) c_{q, +}^{\dagger} \right) ,
\end{align}
where $c_{q, s}$ and $d_{q, s}$
($c_{q, s}^{\dagger}$ and $d_{q, s}^{\dagger}$)
are the fermion anihilation (creation) operators
and $q = \pi (n + 1/2)/(L + \xi)$ with $n = 0, 1, 2, \cdots$.
The linearized quasiparticle dispersion is
\begin{equation}
   \label{eq:qp-e}
  \epsilon_{q\pm} = v_{\rm F} \left( q \pm \frac{\chi}{2(L+\xi)} \right)
                 + \frac{1}{2m} \left( \frac{\chi}{2(L+\xi)} \right)^{2} .
\end{equation}
Thus, the quasiparticles in the 1D system are described by the Hamiltonian
\begin{align}
   \label{eq:Hamil-qp}
  H_{0}
 & = \frac{v_{\rm F}}{4 \pi} \cdot \frac{\chi^{2}}{L+\xi}
              \nonumber \\
 & \hspace{5mm}
  + \sum_{q>0} v_{\rm F} \left(q + \frac{\chi}{2(L+\xi)} \right)
      \left( c_{q, +}^{\dagger} c_{q, +}
              + d_{q, -}^{\dagger}d_{q, -} \right)
              \nonumber \\
 & \hspace{5mm}
  + \sum_{q>0} v_{\rm F} \left(q - \frac{\chi}{2(L+\xi)} \right)
      \left( c_{q, -}^{\dagger} c_{q, -}
              + d_{q, +}^{\dagger}d_{q, +} \right) .
\end{align}
Here, the first term represents the $\chi$-dependent component of
the ground-state energy, arising from the second term in eq.~(\ref{eq:qp-e}).

We define the density operator $\rho_{j s}(x)$ ($j=1,2$) as
$\rho_{j s}(x) = \psi_{j s}^{\dagger}(x) \psi_{j s}(x)$.
Its Fourier transform is given by
\begin{equation}
   \label{eq:dens-op}
  \rho_{j s}(q) = \int_{0}^{L} {\rm d}x {\rm e}^{{\rm i} q x}
                    \rho_{j s}(x) ,
\end{equation}
where $q = \pi (n + 1/2)/(L + \xi)$ with $n = 0, \pm 1, \pm 2, \cdots$.
An important role in our bosonization procedure is played by
the new density operators,
\begin{align}
   \label{eq:dens-op-1}
  \rho_{a}(q) & = {\rm e}^{{\rm i}\xi q/2} \rho_{1 +}(q)
                  - {\rm e}^{-{\rm i}\xi q/2} \rho_{2 -}(-q) , \\
   \label{eq:dens-op-b}
  \rho_{b}(q) & = {\rm e}^{{\rm i}\xi q/2} \rho_{1 -}(q)
                  - {\rm e}^{-{\rm i}\xi q/2} \rho_{2 +}(-q) .
\end{align}
For $q > 0$, they are expressed as
\begin{align}
   \label{eq:dens-a-q+}
 \rho_{a}(q) & = \sum_{q_{1}>0}
       \left( c_{q_{1}+q, +}^{\dagger}c_{q_{1}, +}
            + c_{q_{1}, -}c_{q_{1}+q, -}^{\dagger}
                            \right)
              \nonumber \\
 & \hspace{15mm}
      - {\rm i}\sum_{q>q_{1}>0}
              c_{-q_{1}+q, +}^{\dagger}
              c_{q_{1}, -}^{\dagger} , \\
   \label{eq:dens-a-q-}
 \rho_{a}(-q) & = \sum_{q_{1}>0}
       \left( c_{q_{1}, +}^{\dagger}c_{q_{1}+q, +}
            + c_{q_{1}+q, -}c_{q_{1}, -}^{\dagger}
                            \right)
              \nonumber \\
 & \hspace{15mm}
      + {\rm i}\sum_{q>q_{1}>0}
              c_{q_{1}, -}c_{-q_{1}+q, +} , \\
   \label{eq:dens-b-q+}
 \rho_{b}(q) & = \sum_{q_{1}>0}
       \left( d_{q_{1}+q, -}^{\dagger}d_{q_{1}, -}
            + d_{q_{1}, +}d_{q_{1}+q, +}^{\dagger}
                            \right)
              \nonumber \\
 & \hspace{15mm}
      - {\rm i}\sum_{q>q_{1}>0}
              d_{-q_{1}+q, -}^{\dagger}
              d_{q_{1}, +}^{\dagger} , \\
   \label{eq:dens-b-q-}
 \rho_{b}(-q) & = \sum_{q_{1}>0}
       \left( d_{q_{1}, -}^{\dagger}d_{q_{1}+q, -}
            + d_{q_{1}+q, +}d_{q_{1}, +}^{\dagger}
                            \right)
              \nonumber \\
 & \hspace{15mm}
      + {\rm i}\sum_{q>q_{1}>0}
              d_{q_{1}, +}d_{-q_{1}+q, -} .
\end{align}
We can show that
\begin{align}
   \label{eq:com-dens-a}
  \left[ \rho_{a}(q), \rho_{a}(-q') \right] & =
           - \frac{Lq}{\pi} \delta_{q, q'} , \\
   \label{eq:com-dens-b}
  \left[ \rho_{b}(q), \rho_{b}(-q') \right] & =
           - \frac{Lq}{\pi} \delta_{q, q'} , \\
   \label{eq:com-dens-ab}
  \left[ \rho_{a}(q), \rho_{b}(-q') \right] & = 0 .
\end{align}
The $q=0$ components,
which characterize an excess charge with respect to the ground state,
are given by 
\begin{align}
   \label{eq:dens-a-Z}
  \rho_{a}(q=0) & = N_{a} + \frac{\chi}{2\pi} , \\
   \label{eq:debs-b-Z}
  \rho_{b}(q=0) & = N_{b} + \frac{\chi}{2\pi} ,
\end{align}
where $\chi / (2 \pi)$ results from the shift of
the occupied-state distribution due to the phase difference $\chi$.
We call $N_{a}$ and $N_{b}$ the winding-number operators,
whose eigenvalues are $0, \pm1, \pm2, \cdots .$
We will employ the zero-mode operators $\varphi_{a}$ and $\varphi_{b}$,
which satisfy $[ N_{a}, \varphi_{a} ] = [ N_{b}, \varphi_{b} ] = {\rm i}$
and $[ N_{a}, \varphi_{b} ] = [ N_{b}, \varphi_{a} ] = 0$.

Let us consider the commutation relations between the density operators
with $q \neq 0$
and the field operators presented in eqs.~(\ref{eq:psi-1+})-(\ref{eq:psi-2-}).
We find that
\begin{align}
   \label{eq:com-a-p1}
   \left[ \rho_{a}(q), \psi_{1+}(x) \right]
 & = - {\rm e}^{{\rm i}q(x+\xi /2)}\psi_{1+}(x) , \\
   \label{eq:com-a-p2}
   \left[ \rho_{a}(q), \psi_{2-}(x) \right]
 & =   {\rm e}^{-{\rm i}q(x+\xi /2)}\psi_{2-}(x) , \\
   \label{eq:com-b-p1}
   \left[ \rho_{b}(q), \psi_{1-}(x) \right]
 & = - {\rm e}^{{\rm i}q(x+\xi /2)}\psi_{1-}(x) , \\
   \label{eq:com-b-p2}
   \left[ \rho_{b}(q), \psi_{2+}(x) \right]
 & =   {\rm e}^{-{\rm i}q(x+\xi /2)}\psi_{2+}(x) , \\
   \label{eq:com-ab-p12}
   \left[ \rho_{a}(q), \psi_{1-}(x) \right]
 & =  \left[ \rho_{a}(q), \psi_{2+}(x) \right]
   =  \left[ \rho_{b}(q), \psi_{1+}(x) \right]
              \nonumber \\
 & =  \left[ \rho_{b}(q), \psi_{2-}(x) \right] = 0 .
\end{align}
Noting eqs.~(\ref{eq:com-a-p1})-(\ref{eq:com-ab-p12}),
we express the field operators as~\cite{rf:32}
\begin{align}
   \label{eq:psi1-dens}
  \psi_{1s}(x) & = \frac{1}{\sqrt{2 \pi \alpha}}
       {\rm e}^{{\rm i}k_{\rm F}x+{\rm i}\theta_{1s}(x)} , \\
   \label{eq:psi2-dens}
  \psi_{2s}(x) & = \frac{1}{\sqrt{2 \pi \alpha}}
       {\rm e}^{-{\rm i}k_{\rm F}x+{\rm i}\theta_{2s}(x)} ,
\end{align}
where
\begin{align}
   \label{eq:theta1+}
  \theta_{1+}(x)
 & = \theta_{1+}^{\rm Z}
       + {\rm i}\frac{\pi}{L} \sum_{q>0} \frac{{\rm e}^{-\alpha q/2}}{q}
          \Big( {\rm e}^{-{\rm i}q(x+\xi /2)} \rho_{a}(q)
              \nonumber \\
 & \hspace{25mm}
                   - {\rm e}^{{\rm i}q(x+\xi /2)} \rho_{a}(-q) \Big) , \\
   \label{eq:theta1-}
 \theta_{1-}(x)
 & = \theta_{1-}^{\rm Z}
       + {\rm i}\frac{\pi}{L} \sum_{q>0} \frac{{\rm e}^{-\alpha q/2}}{q}
          \Big( {\rm e}^{-{\rm i}q(x+\xi /2)} \rho_{b}(q)
              \nonumber \\
 & \hspace{25mm}
                   - {\rm e}^{{\rm i}q(x+\xi /2)} \rho_{b}(-q) \Big) , \\
   \label{eq:theta2+}
 \theta_{2+}(x)
 & = \theta_{2+}^{\rm Z}
       + {\rm i}\frac{\pi}{L} \sum_{q>0} \frac{{\rm e}^{-\alpha q/2}}{q}
          \Big( - {\rm e}^{{\rm i}q(x+\xi /2)} \rho_{b}(q)
              \nonumber \\
 & \hspace{25mm}
                   + {\rm e}^{-{\rm i}q(x+\xi /2)} \rho_{b}(-q) \Big) , \\
   \label{eq:theta2-}
 \theta_{2-}(x)
 & = \theta_{2-}^{\rm Z}
       + {\rm i}\frac{\pi}{L} \sum_{q>0} \frac{{\rm e}^{-\alpha q/2}}{q}
          \Big( - {\rm e}^{{\rm i}q(x+\xi /2)} \rho_{a}(q)
              \nonumber \\
 & \hspace{25mm}
                   + {\rm e}^{-{\rm i}q(x+\xi /2)} \rho_{a}(-q) \Big) .
\end{align}
Here, $\alpha$ is a positive infinitesimal and
$\theta_{j s}^{\rm Z}$ ($j=1, 2$) represents the zero-mode component
to be determined.
The phase field $\theta_{j s}$ is related
with the density operator $\rho_{j s}$,
\begin{equation}
   \label{eq:theta-rho}
  \rho_{js}(x) = \frac{1}{2\pi} \frac{\partial \theta_{j s}}{\partial x} .
\end{equation}
For example, we find that
\begin{align}
   \label{eq:rho1+x}
  \rho_{1+}(x)
  & =   \rho_{1+}^{\rm Z}(x)
      + \frac{1}{2L} \sum_{q>0}
        \Big( {\rm e}^{-{\rm i}q(x+\xi /2)} \rho_{a}(q)
              \nonumber \\
 & \hspace{27mm}
                + {\rm e}^{{\rm i}q(x+\xi /2)} \rho_{a}(-q) \Big) , \\
   \label{eq:rho2-x}
  \rho_{2-}(x)
  & =   \rho_{2-}^{\rm Z}(x)
      + \frac{1}{2L} \sum_{q>0}
        \Big( {\rm e}^{{\rm i}q(x+\xi /2)} \rho_{a}(q)
              \nonumber \\
 & \hspace{27mm}
                + {\rm e}^{-{\rm i}q(x+\xi /2)} \rho_{a}(-q) \Big) ,
\end{align}
where
\begin{equation}
   \label{eq:rho+Zx}
 \rho_{j s}^{\rm Z}(x)
    = \frac{1}{2\pi}
       \frac{\partial \theta_{js}^{\rm Z}}{\partial x} .
\end{equation}
From eqs.~(\ref{eq:rho1+x}) and (\ref{eq:rho2-x}), we naturally find that
\begin{equation}
   \label{eq:rho1+Zx2}
 \rho_{1+}^{\rm Z}(x) = \rho_{2-}^{\rm Z}(x)
    = \frac{\pi}{L} \left( N_{a} + \frac{\chi}{2\pi} \right) .
\end{equation}
We determin $\theta_{j s}^{\rm Z}$ so as to ensure the fermion commutation
relation between $\psi_{j\pm}$ and $\psi_{j'\pm}$
and that between $\psi_{j\pm}$ and $\psi_{j'\pm}^{\dagger}$ ($j, j' = 1, 2$).
The result is
\begin{align}
   \label{eq:theta1+2}
  \theta_{1 +}^{\rm Z}(x)
 & = \varphi_{a}
      + \frac{\pi}{L} \left( N_{a} + \frac{\chi}{2\pi} \right)
          \left( x + \frac{\xi}{2} \right)
              \nonumber \\
 & \hspace{35mm}
      + \frac{\pi}{2} \left(N_{a} -N_{b} \right) , \\
   \label{eq:theta2-2}
  \theta_{2 -}^{\rm Z}(x)
 & = - \varphi_{a}
      + \frac{\pi}{L} \left( N_{a} + \frac{\chi}{2\pi} \right)
          \left( x + \frac{\xi}{2} \right)
              \nonumber \\
 & \hspace{35mm}
      + \frac{\pi}{2} \left(N_{a} -N_{b} \right) , \\
   \label{eq:theta1-2}
  \theta_{1 -}^{\rm Z}(x)
 & = \varphi_{b}
      + \frac{\pi}{L} \left( N_{b} + \frac{\chi}{2\pi} \right)
          \left( x + \frac{\xi}{2} \right)
              \nonumber \\
 & \hspace{35mm}
      - \frac{\pi}{2} \left(N_{a} -N_{b} \right) , \\
   \label{eq:theta2+2}
  \theta_{2 +}^{\rm Z}(x)
 & = - \varphi_{b}
      + \frac{\pi}{L} \left( N_{b} + \frac{\chi}{2\pi} \right)
          \left( x + \frac{\xi}{2} \right)
              \nonumber \\
 & \hspace{35mm}
      - \frac{\pi}{2} \left(N_{a} -N_{b} \right) , 
\end{align}
where $\varphi_{a}$ and $\varphi_{b}$ are the zero-mode operators
mentioned above.
Strictly speaking, if we need to ensure the fermion commutation
relation between $\psi_{j\pm}$ and $\psi_{j'\mp}$
and that between $\psi_{j\pm}$ and $\psi_{j'\mp}^{\dagger}$,
Majorana fermions must be introduced in
eqs.~(\ref{eq:psi1-dens}) and (\ref{eq:psi2-dens}).
However, since the Majorana fermions do not play an essential role
in our argument, we neglect them in the following.
For actual calculations, it is convenient to use
the new phase fields:~\cite{rf:31}
\begin{align}
   \label{eq:Theta+A}
  \Theta_{+}(x) & = \frac{1}{2}
     \Big( \theta_{1 +}(x) + \theta_{1 -}(x)
          - \theta_{2 +}(x) - \theta_{2 -}(x) \Big) , \\
   \label{eq:Theta-A}
  \Theta_{-}(x) & = \frac{1}{2}
     \Big( \theta_{1 +}(x) + \theta_{1 -}(x)
          + \theta_{2 +}(x) + \theta_{2 -}(x) \Big) , \\
   \label{eq:Phi+A}
  \Phi_{+}(x) & = \frac{1}{2}
     \Big( \theta_{1 +}(x) - \theta_{1 -}(x)
          - \theta_{2 +}(x) + \theta_{2 -}(x) \Big) , \\
   \label{eq:Phi-A}
  \Phi_{-}(x) & = \frac{1}{2}
     \Big( \theta_{1 +}(x) - \theta_{1 -}(x)
          + \theta_{2 +}(x) - \theta_{2 -}(x) \Big) .
\end{align}
To express these phase fields compactly, 
we define $\varphi_{\rho}$, $\varphi_{\sigma}$, $J$ and $M$ as
\begin{align}
  \varphi_{\rho} & = \varphi_{a} + \varphi_{b} , \\
  \varphi_{\sigma} & = \varphi_{a} - \varphi_{b} , \\
  J & = N_{a} + N_{b} , \\
  M & = N_{a} - N_{b} .
\end{align}
It is easy to show that $[J, \varphi_{\rho}] =
[M, \varphi_{\sigma}] = 2 {\rm i}$ and
$[J, \varphi_{\sigma}] = [M, \varphi_{\rho}] = 0$.
Substituting eqs.~(\ref{eq:theta1+2})-(\ref{eq:theta2+2})
into the above expressions and using these operators,
we find that
\begin{align}
   \label{eq:Theta+B}
 \Theta_{+}(x) & = \varphi_{\rho} + \theta_{+}(x) , \\
   \label{eq:Theta-B}
 \Theta_{-}(x) & = \frac{\pi}{L}\left(J + \frac{\chi}{\pi} \right)
                                     \left(x + \frac{\xi}{2} \right)
                     + \theta_{-}(x) , \\
   \label{eq:Phi+B}
 \Phi_{+}(x) & = \frac{\pi}{L} M \left(x + L + \frac{\xi}{2} \right)
                     + \phi_{+}(x) , \\
   \label{eq:Phi-B}
 \Phi_{-}(x) & = \varphi_{\sigma} + \phi_{-}(x) ,
\end{align}
where
\begin{align}
   \label{eq:theta+C}
  \theta_{+}(x)
 & = {\rm i} \frac{\pi}{L} \sum_{q>0} \frac{{\rm e}^{-\alpha q/2}}{q}
          \cos q \left(x + \frac{\xi}{2} \right)
              \nonumber \\
 & \hspace{3mm} \times
           \left( \rho_{a}(q) - \rho_{a}(-q)
                   + \rho_{b}(q) - \rho_{b}(-q) \right) , \\
   \label{eq:theta-C}
  \theta_{-}(x)
 & = \frac{\pi}{L} \sum_{q>0} \frac{{\rm e}^{-\alpha q/2}}{q}
          \sin q \left(x + \frac{\xi}{2} \right)
              \nonumber \\
 & \hspace{3mm} \times
           \left( \rho_{a}(q) + \rho_{a}(-q)
                   + \rho_{b}(q) + \rho_{b}(-q) \right) , \\
   \label{eq:phi+C}
  \phi_{+}(x)
 & = \frac{\pi}{L} \sum_{q>0} \frac{{\rm e}^{-\alpha q/2}}{q}
          \sin q \left(x + \frac{\xi}{2} \right)
              \nonumber \\
 & \hspace{3mm} \times
           \left( \rho_{a}(q) + \rho_{a}(-q)
                   - \rho_{b}(q) - \rho_{b}(-q) \right) , \\
   \label{eq:phi-C}
 \phi_{-}(x)
 & = {\rm i} \frac{\pi}{L} \sum_{q>0} \frac{{\rm e}^{-\alpha q/2}}{q}
          \cos q \left(x + \frac{\xi}{2} \right)
              \nonumber \\
 & \hspace{3mm} \times
           \left( \rho_{a}(q) - \rho_{a}(-q)
                   - \rho_{b}(q) + \rho_{b}(-q) \right) .
\end{align}
Using the phase fields, we rewrite
eqs.~(\ref{eq:psi1-dens}) and (\ref{eq:psi2-dens}) as~\cite{rf:31, rf:32}
\begin{align}
   \label{eq:psi1-A}
  \psi_{1s}(x) & = \frac{1}{\sqrt{2 \pi \alpha}}
       {\rm e}^{{\rm i}k_{\rm F}x + \frac{\rm i}{2}
         \left( \Theta_{+}(x) + \Theta_{-}(x)
                   + s \Phi_{+}(x) + s \Phi_{-}(x) \right) } , \\
   \label{eq:psi2-A}
  \psi_{2s}(x) & = \frac{1}{\sqrt{2 \pi \alpha}}
       {\rm e}^{-{\rm i}k_{\rm F}x + \frac{\rm i}{2}
         \left( -\Theta_{+}(x) + \Theta_{-}(x)
                   - s \Phi_{+}(x) + s \Phi_{-}(x) \right) } .
\end{align}

We note that the density operators with $q \neq 0$
and $H_{0}$ satisfy following relations
\begin{align}
   \label{eq:com-a-H}
   \left[ \rho_{a}(q), H_{0} \right]
 & = - v_{\rm F} q \rho_{a}(q) , \\
   \label{eq:com-b-H}
   \left[ \rho_{b}(q), H_{0} \right]
 & = - v_{\rm F} q \rho_{b}(q) .
\end{align}
We separate the nonzero-mode component $H_{0}^{\rm NZ}$ from $H_{0}$.
From eqs.~(\ref{eq:com-a-H}) and (\ref{eq:com-b-H}) we find that
\begin{equation}
   \label{eq:Hamil-0-NZ}
 H_{0}^{\rm NZ} = \frac{\pi v_{\rm F}}{L}
            \sum_{q>0} \left( \rho_{a}(q)\rho_{a}(-q)
                        + \rho_{b}(q)\rho_{b}(-q) \right) .
\end{equation}
Next we consider the zero-mode component $H_{0}^{\rm Z}$.
The energy cost due to the excess charge $N_{a} ( > 0 )$ is obtained as
\begin{align}
   \label{eq:Delta-E-a}
  \Delta E(N_{a})
 & = \sum_{n=0}^{N_{a} - 1}
       v_{\rm F} \left(\frac{\pi(n + 1/2)}{L + \xi}
                      + \frac{\chi}{2(L + \xi)} \right)
              \nonumber \\
 & \approx
    \frac{v_{\rm F}\pi}{2L} N_{a}^{2}
     + \frac{v_{\rm F}\chi}{2L} N_{a} .
\end{align}
This expression also holds for negative $N_{a}$.
Similarly, we obtain the energy cost due to $N_{b}$.
Adding $\Delta E(N_{a})$, $\Delta E(N_{b})$ and the $\chi$-dependent
part of the ground-state enrgy $(v_{\rm F}/4\pi)(\chi^{2}/L)$,
we obtain
\begin{equation}
   \label{eq:Hamil-0-Z}
 H_{0}^{\rm Z} = \frac{v_{\rm F}\pi}{2L}
        \left(  \left( N_{a}+\frac{\chi}{2\pi} \right)^{2}
              + \left( N_{b}+\frac{\chi}{2\pi} \right)^{2} \right) .
\end{equation}

We have expressed the field operators and the Hamiltonian
in terms of the density operators.
In order to simplify the expressions,
we introduce operators for $q > 0$:
\begin{align}
   \label{eq:boson-a}
 \alpha_{q} & = \sqrt{\frac{\pi}{2qL}}
              \left(\rho_{a}(-q) + \rho_{b}(-q) \right) , \\
   \label{eq:boson-a-d}
 \alpha_{q}^{\dagger} & = \sqrt{\frac{\pi}{2qL}}
              \left(\rho_{a}(q) + \rho_{b}(q) \right) , \\
   \label{eq:boson-b}
 \beta_{q} & = \sqrt{\frac{\pi}{2qL}}
              \left(\rho_{a}(-q) - \rho_{b}(-q) \right) , \\
   \label{eq:boson-b-d}
 \beta_{q}^{\dagger} & = \sqrt{\frac{\pi}{2qL}}
              \left(\rho_{a}(q) - \rho_{b}(q) \right) .
\end{align}
They satisfy the boson commutation relation.
Using the boson operators, we rewrite the phase fields defined
in eqs.~(\ref{eq:theta+C})-(\ref{eq:phi+C}) as
\begin{align}
   \label{eq:theta+boson}
 \theta_{+}(x) & = {\rm i} \sum_{q>0} \sqrt{\frac{2\pi}{qL}}
        {\rm e}^{-\alpha q/2} \cos q \left(x + \frac{\xi}{2} \right)
          \left(\alpha_{q}^{\dagger} - \alpha_{q} \right), \\
   \label{eq:theta-boson}
 \theta_{-}(x) & = \sum_{q>0} \sqrt{\frac{2\pi}{qL}}
       {\rm e}^{-\alpha q/2} \sin q \left(x + \frac{\xi}{2} \right)
          \left(\alpha_{q}^{\dagger} + \alpha_{q} \right), \\
   \label{eq:phi+boson}
 \phi_{+}(x) & = \sum_{q>0} \sqrt{\frac{2\pi}{qL}}
        {\rm e}^{-\alpha q/2} \sin q \left(x + \frac{\xi}{2} \right)
          \left(\beta_{q}^{\dagger} + \beta_{q} \right), \\
   \label{eq:phi-boson}
 \phi_{-}(x) & = {\rm i} \sum_{q>0} \sqrt{\frac{2\pi}{qL}}
        {\rm e}^{-\alpha q/2} \cos q \left(x + \frac{\xi}{2} \right)
          \left(\beta_{q}^{\dagger} - \beta_{q} \right) .
\end{align}
Similarly, the Hamiltonian is expressed as
\begin{equation}
   \label{eq:Hamil-boson}
 H_{0} = \frac{v_{\rm F}\pi}{4L}
           \left(  \left( J + \frac{\chi}{\pi} \right)^{2}
                                  + M^{2} \right)
   + \sum_{q > 0} v_{\rm F} q \left(  \alpha_{q}^{\dagger}\alpha_{q} 
                                + \beta_{q}^{\dagger}\beta_{q} \right) .
\end{equation}
Note that since $J = N_{a} + N_{b}$ and $M = N_{a} - N_{b}$,
then $J$ must be even (odd) if $M$ is even (odd).
That is, $J + M = {\rm even}$.

We take account of electron-electron interactions in the following.
We employ a model interaction Hamiltonian $H_{\rm int}$ given as follows:
\begin{align}
   \label{eq:Hamil-int}
   H_{\rm int}
 & = 2\pi v_{\rm F} \int_{0}^{L} {\rm d}x \sum_{s, s'}
         \left(g_{2} \delta_{s, s'} + g'_{2} \delta_{s, -s'} \right)
            \rho_{1s}(x) \rho_{2s'}(x)
              \nonumber \\
 & \hspace{5mm}
   + \pi v_{\rm F} \int_{0}^{L} {\rm d}x \sum_{s, s'}
         \left(g_{4} \delta_{s, s'} + g'_{4} \delta_{s, -s'} \right)
              \nonumber \\
 & \hspace{25mm} \times
            \left( \rho_{1s}(x) \rho_{1s'}(x)
                      + \rho_{2s}(x) \rho_{2s'}(x) \right) .
\end{align}
In terms of $\rho_{a}(q)$ and $\rho_{b}(q)$,
the interaction Hamiltonian is rewritten as
\begin{align}
   \label{eq:Hamil-int2}
   H_{\rm int}
 & = - \frac{\pi v_{\rm F} g_{2}}{L} \sum_{q} \rho_{a}(q) \rho_{b}(q)
              \nonumber \\
 & \hspace{5mm}
     + \frac{\pi v_{\rm F}g'_{2}}{2L} \sum_{q}
        \left( \rho_{a}(q) \rho_{a}(q) + \rho_{b}(q) \rho_{b}(q) \right)
              \nonumber \\
 & \hspace{5mm}
     + \frac{\pi v_{\rm F}g_{4}}{2L} \sum_{q}
        \left( \rho_{a}(q) \rho_{a}(-q) + \rho_{b}(q) \rho_{b}(-q) \right)
              \nonumber \\
 & \hspace{5mm}
     + \frac{\pi v_{\rm F} g'_{4}}{L} \sum_{q} \rho_{a}(q) \rho_{b}(-q) .
\end{align}
We decompose $H_{\rm int}$ into the nonzero-mode component
$H_{\rm int}^{\rm NZ}$ and the zero-mode component $H_{\rm int}^{\rm Z}$,
\begin{align}
   \label{eq:Hamil-int-NZ}
 H_{\rm int}^{\rm NZ}
 & = - \left(g_{2} + g'_{2} \right) \sum_{q > 0} \frac{v_{\rm F}q}{2}
                 \left( \alpha_{q}^{\dagger}\alpha_{q}^{\dagger}
                              + \alpha_{q}\alpha_{q} \right)
              \nonumber \\
 & \hspace{3mm}
     + \left(g_{2} - g'_{2} \right) \sum_{q > 0} \frac{v_{\rm F}q}{2}
                 \left( \beta_{q}^{\dagger}\beta_{q}^{\dagger}
                              + \beta_{q}\beta_{q} \right)
              \nonumber \\
 & \hspace{3mm}
    + \left(g_{4} + g'_{4} \right) \sum_{q > 0} v_{\rm F}q
                     \alpha_{q}^{\dagger}\alpha_{q}
    + \left(g_{4} - g'_{4} \right) \sum_{q > 0} v_{\rm F}q
                     \beta_{q}^{\dagger}\beta_{q} , \\
   \label{eq:Hamil-int-Z}
 H_{\rm int}^{\rm Z}
 & = - \frac{\pi v_{\rm F}}{4L} \left(g_{2} + g'_{2} - g_{4} - g'_{4} \right)
          \left( J + \frac{\chi}{\pi} \right)^{2}
              \nonumber \\
 & \hspace{3mm}
     + \frac{\pi v_{\rm F}}{4L} \left(g_{2} - g'_{2} + g_{4} - g'_{4} \right)
                 M^{2} .
\end{align}
The nonzero-mode part
$H^{\rm NZ} \equiv H_{0}^{\rm NZ} + H_{\rm int}^{\rm NZ}$ is diagonalized as
\begin{equation}
   \label{eq:Hamil-NZ-A}
 H^{\rm NZ} = \sum_{q > 0} \left( v_{\rho} q a_{q}^{\dagger}a_{q}
                               + v_{\sigma} q b_{q}^{\dagger}b_{q}\right) ,
\end{equation}
where
\begin{equation}
   \label{eq:v-rho}
  v_{\rho (\sigma)} = v_{\rm F}
     \sqrt{ \left( 1 + g_{4} \pm g'_{4} \right)^{2}
                           -\left( g_{2} \pm g'_{2} \right)^{2} } .
\end{equation}
Here, the new boson operators are given by
\begin{align}
   \label{eq:boson-a-new}
  a_{q} & = \alpha_{q} \cosh \lambda_{\rho}
                - \alpha_{q}^{\dagger} \sinh \lambda_{\rho} , \\
   \label{eq:boson-b-new}
  b_{q} & = \beta_{q} \cosh \lambda_{\sigma}
                + \beta_{q}^{\dagger} \sinh \lambda_{\sigma} ,
\end{align}
with
\begin{equation}
   \label{eq:diago-tanh}
  \tanh \left( 2\lambda_{\rho (\sigma)} \right)
      =\frac{g_{2} \pm g'_{2}}{1 + g_{4} \pm g'_{4}} .
\end{equation}
The zero-mode part $H^{\rm Z} \equiv H_{0}^{\rm Z} + H_{\rm int}^{\rm Z}$
is expressed as
\begin{equation}
   \label{eq:Hamil-Z-A}
  H^{\rm Z} = \frac{\pi}{4L}
      \left( v_{\rho} K_{\rho} \left( J + \frac{\chi}{\pi} \right)^{2}
            + \frac{v_{\sigma}}{K_{\sigma}} M^{2} \right) ,
\end{equation}
where $K_{\rho (\sigma)}$ is the correlation exponent
\begin{equation}
   \label{eq:K-rho}
  K_{\rho (\sigma)} =
    \sqrt{\frac{ 1 + \left( g_{4} \pm g'_{4} \right)
                   - \left( g_{2} \pm g'_{2} \right)}
               { 1 + \left( g_{4} \pm g'_{4} \right)
                   + \left( g_{2} \pm g'_{2} \right)} } .
\end{equation}
Consequently, the total Hamiltonian is
\begin{align}
   \label{eq:Hamil-total}
  H & = \frac{\pi}{4L}
        \left( v_{\rho} K_{\rho} \left( J + \frac{\chi}{\pi} \right)^{2}
            + \frac{v_{\sigma}}{K_{\sigma}} M^{2} \right) 
              \nonumber \\
    & \hspace{5mm}
      + \sum_{q > 0} \left( v_{\rho} q a_{q}^{\dagger}a_{q}
                               + v_{\sigma} q b_{q}^{\dagger}b_{q}\right) .
\end{align}
Finally, we rewrite $\theta_{\pm}$ and $\phi_{\pm}$ in terms of
$a_{q}$, $a_{q}^{\dagger}$, $b_{q}$ and $b_{q}^{\dagger}$ as
\begin{align}
   \label{eq:theta+final}
  \theta_{+}(x)
 & = {\rm i} \sqrt{K_{\rho}}
        \sum_{q>0} \sqrt{\frac{2\pi}{qL}}
        {\rm e}^{-\alpha q/2} \cos q \left(x + \frac{\xi}{2} \right)
          \left(a_{q}^{\dagger} - a_{q} \right), \\
   \label{eq:theta-final}
  \theta_{-}(x)
 & = \frac{1}{\sqrt{K_{\rho}}}
        \sum_{q>0} \sqrt{\frac{2\pi}{qL}}
        {\rm e}^{-\alpha q/2} \sin q \left(x + \frac{\xi}{2} \right)
          \left(a_{q}^{\dagger} + a_{q} \right), \\
   \label{eq:phi+final}
  \phi_{+}(x)
 & = \sqrt{K_{\sigma}}
        \sum_{q>0} \sqrt{\frac{2\pi}{qL}}
        {\rm e}^{-\alpha q/2} \sin q \left(x + \frac{\xi}{2} \right)
          \left(b_{q}^{\dagger} + b_{q} \right), \\
   \label{eq:phi-final}
  \phi_{-}(x)
 & = {\rm i} \frac{1}{K_{\sigma}}
        \sum_{q>0} \sqrt{\frac{2\pi}{qL}}
        {\rm e}^{-\alpha q/2} \cos q \left(x + \frac{\xi}{2} \right)
          \left(b_{q}^{\dagger} - b_{q} \right) .
\end{align}

\section{Derivation of $S_{\rm eff}^{\rm NZ}$}

Let us consider $Z^{\rm NZ}$ defined as
\begin{align}
   \label{eq:pf-NZ1}
  Z^{\rm NZ}
 & = \int \prod_{q>0}
           \mathcal{D}a_{q}^{\dagger} \mathcal{D}a_{q}
           \mathcal{D}b_{q}^{\dagger} \mathcal{D}b_{q}
              \nonumber \\
 & \hspace{5mm} \times
     \exp \left( 
                 - S^{\rm NZ} \bigl[ \{a_{q}^{\dagger}, a_{q},
                                  b_{q}^{\dagger}, b_{q}\} \bigr]
                 - S^{\rm B} \bigl[ \{a_{q}^{\dagger}, a_{q},
                                  b_{q}^{\dagger}, b_{q}\} \bigr]
          \right) .
\end{align}
We can rewrite $Z^{\rm NZ}$ as
\begin{align}
   \label{eq:pf-NZ2}
   Z^{\rm NZ}
 & = \int \prod_{q>0}
           \mathcal{D}a_{q}^{\dagger} \mathcal{D}a_{q}
           \mathcal{D}b_{q}^{\dagger} \mathcal{D}b_{q}
     \int \mathcal{D}\bar{\theta}\mathcal{D}\tilde{\theta}
          \mathcal{D}\bar{\phi}\mathcal{D}\tilde{\phi}
              \nonumber \\
 & \hspace{5mm} \times
     \delta \left( \bar{\theta} - \frac{1}{2}
                         \left( \theta_{+}(L) + \theta_{+}(0) \right)
           \right)
              \nonumber \\
 & \hspace{5mm} \times
     \delta \left( \tilde{\theta} - 
                         \left( \theta_{+}(L) - \theta_{+}(0) \right)
           \right)
              \nonumber \\
 & \hspace{5mm} \times
     \delta \left( \bar{\phi} - \frac{1}{2}
                         \left( \phi_{+}(L) + \phi_{+}(0) \right)
           \right)
              \nonumber \\
 & \hspace{5mm} \times
     \delta \left( \tilde{\phi} - 
                         \left( \phi_{+}(L) - \phi_{+}(0) \right)
           \right)
              \nonumber \\
 & \hspace{5mm} \times
     \exp \left( 
                 - S^{\rm NZ} \bigl[ \{a_{q}^{\dagger}, a_{q},
                                    b_{q}^{\dagger}, b_{q}\} \bigr]
                 - S^{\rm B} \bigl[ \bar{\theta}, \tilde{\theta},
                                    \bar{\phi}, \tilde{\phi} \bigr]
          \right)
              \nonumber \\
 & \propto
    \int \mathcal{D}\bar{\theta}\mathcal{D}\tilde{\theta}
         \mathcal{D}\bar{\phi}\mathcal{D}\tilde{\phi}
    \exp \left( - S^{\rm B} \bigl[ \bar{\theta}, \tilde{\theta},
                                    \bar{\phi}, \tilde{\phi} \bigr]
         \right)
              \nonumber \\
 & \hspace{5mm} \times
    \int \mathcal{D}\bar{\lambda_{\rho}}\mathcal{D}\tilde{\lambda_{\rho}}
         \mathcal{D}\bar{\lambda_{\sigma}}\mathcal{D}\tilde{\lambda_{\sigma}}
    \int \prod_{q>0}
           \mathcal{D}a_{q}^{\dagger} \mathcal{D}a_{q}
           \mathcal{D}b_{q}^{\dagger} \mathcal{D}b_{q}
              \nonumber \\
 & \hspace{5mm} \times
    \exp \bigg( 
               - S^{\rm NZ} \bigl[ \{a_{q}^{\dagger}, a_{q},
                                      b_{q}^{\dagger}, b_{q}\} \bigr]
              \nonumber \\
 & \hspace{7mm}
               + {\rm i} \mathcal{P}\bigl[ \{a_{q}^{\dagger}, a_{q},
                                      b_{q}^{\dagger}, b_{q}\},
                                      \bar{\theta}, \tilde{\theta},
                                      \bar{\phi}, \tilde{\phi},
                                      \bar{\lambda_{\rho}},
                                      \tilde{\lambda_{\rho}},
                                      \bar{\lambda_{\sigma}},
                                      \tilde{\lambda_{\sigma}}
                                      \bigr]
         \bigg) .
\end{align}
where
\begin{align}
   \label{eq:P-in-pf}
  \mathcal{P}
 & = \int {\rm d}\tau 
       \Bigg[  \bar{\lambda_{\rho}}(\tau)
                 \biggl( \bar{\theta}(\tau)
                        - \frac{1}{2} \left( \theta_{+}(L, \tau)
                                  + \theta_{+}(0, \tau) \right)
                 \biggr)
              \nonumber \\
 & \hspace{5mm}
             + \tilde{\lambda_{\rho}}(\tau)
                 \biggl( \tilde{\theta}(\tau)
                        - \left( \theta_{+}(L, \tau) - \theta_{+}(0, \tau)
                          \right)
                 \biggr)
              \nonumber \\
 & \hspace{5mm}
             + \bar{\lambda_{\sigma}}(\tau)
                 \biggl( \bar{\phi}(\tau)
                        - \frac{1}{2} \left( \phi_{+}(L,\tau)
                            + \phi_{+}(0, \tau) \right)
                 \biggr)
              \nonumber \\
 & \hspace{5mm}
             + \tilde{\lambda_{\sigma}}(\tau)
                 \biggl( \tilde{\phi}(\tau) - \left( \phi_{+}(L, \tau)
                                         - \phi_{+}(0, \tau) \right)
                 \biggr)
       \Biggr] .
\end{align}
Note that $\theta_{+}(L)$ and $\theta_{+}(0)$
in eqs.~(\ref{eq:pf-NZ2}) and (\ref{eq:P-in-pf}) are functions of
$\{a_{q}^{\dagger}, a_{q} \}$ as shown in eq.~(\ref{eq:f-theta-+}).
Similarly, $\phi_{+}(L)$ and $\phi_{+}(0)$ are functions of
$\{b_{q}^{\dagger}, b_{q} \}$.
Integration over $a_{q}$, $a_{q}^{\dagger}$, $b_{q}$ and $b_{q}^{\dagger}$
yields
\begin{align}
   \label{eq:pf-NZ3}
& \int \mathcal{D}\bar{\lambda_{\rho}}\mathcal{D}\tilde{\lambda_{\rho}}
         \mathcal{D}\bar{\lambda_{\sigma}}\mathcal{D}\tilde{\lambda_{\sigma}}
  \int \prod_{q>0}
         \mathcal{D}a_{q}^{\dagger} \mathcal{D}a_{q}
         \mathcal{D}b_{q}^{\dagger} \mathcal{D}b_{q}
        \nonumber \\
& \hspace{5mm} \times
  \exp \bigg( 
          - S^{\rm NZ} \bigl[ \{a_{q}^{\dagger}, a_{q},
                                    b_{q}^{\dagger}, b_{q}\} \bigr]
              \nonumber \\
 & \hspace{10mm} \times
          + {\rm i} \mathcal{P}\bigl[ \{a_{q}^{\dagger}, a_{q},
                                        b_{q}^{\dagger}, b_{q}\},
                                      \bar{\theta}, \tilde{\theta},
                                      \bar{\phi}, \tilde{\phi},
                                      \bar{\lambda_{\rho}},
                                      \tilde{\lambda_{\rho}},
                                      \bar{\lambda_{\sigma}},
                                      \tilde{\lambda_{\sigma}}
                                      \bigr]
       \bigg)
        \nonumber \\
& = \int \mathcal{D}\bar{\lambda_{\rho}}\mathcal{D}\tilde{\lambda_{\rho}}
         \mathcal{D}\bar{\lambda_{\sigma}}\mathcal{D}\tilde{\lambda_{\sigma}}
              \nonumber \\
 & \hspace{3mm} \times
    \exp \Bigg( - \frac{\pi K_{\rho}}{2 \beta} \sum_{\omega}
                 \bar{J}_{\rho}(\omega)
                 \bar{\lambda}_{\rho}(\omega)\bar{\lambda}_{\rho}(-\omega)
              \nonumber \\
 & \hspace{37mm}
            - {\rm i}\frac{1}{\beta}\sum_{\omega}
                      \bar{\theta}(\omega)\bar{\lambda}_{\rho}(-\omega)
        \nonumber \\
 & \hspace{15mm}
            - \frac{2\pi K_{\rho}}{\beta}\sum_{\omega}
                 \tilde{J}_{\rho}(\omega)
                 \tilde{\lambda}_{\rho}(\omega)
                               \tilde{\lambda}_{\rho}(-\omega)
              \nonumber \\
 & \hspace{37mm}
             - {\rm i}\frac{1}{\beta}\sum_{\omega}
                      \tilde{\theta}(\omega)\tilde{\lambda}_{\rho}(-\omega)
        \nonumber \\
 & \hspace{15mm}
            - \frac{\pi K_{\sigma}}{2 \beta} \sum_{\omega}
                 \bar{J}_{\sigma}(\omega)
                      \bar{\lambda}(\omega)_{\sigma}
                               \bar{\lambda}_{\sigma}(-\omega)
              \nonumber \\
 & \hspace{37mm}
             - {\rm i}\frac{1}{\beta}\sum_{\omega}
                      \bar{\phi}(\omega)\bar{\lambda}_{\sigma}(-\omega)
        \nonumber \\
 & \hspace{15mm}
            - \frac{2\pi K_{\sigma}}{\beta}\sum_{\omega}
                 \tilde{J}_{\sigma}(\omega)
                      \tilde{\lambda}_{\sigma}(\omega)
                               \tilde{\lambda}_{\sigma}(-\omega)
              \nonumber \\
 & \hspace{37mm}
            - {\rm i}\frac{1}{\beta}\sum_{\omega}
                      \tilde{\phi}(\omega)\tilde{\lambda}_{\sigma}(-\omega)
         \Bigg) ,
\end{align}
where
\begin{align}
   \label{eq:bar-J-rho-A}
 \bar{J}_{\rho}(\omega)
& = \frac{1}{2 \omega \sinh \left( \frac{L \omega}{v_{\rho}} \right)}
    \Bigg\{  \cosh \left( \frac{L \omega}{v_{\rho}} \right)
              \nonumber \\
& \hspace{-2mm}
         + \cosh \left( \frac{(L-\xi) \omega}{v_{\rho}} \right)
         + \cosh \left( \frac{\xi \omega}{v_{\rho}} \right) +1 \Bigg\}
            -\frac{2v_{\rho}}{L \omega^{2}} , \\
   \label{eq:tilde-J-rho-A}
 \tilde{J}_{\rho}(\omega)
& = \frac{1}{2 \omega \sinh \left( \frac{L \omega}{v_{\rho}} \right)}
    \Bigg\{  \cosh \left( \frac{L \omega}{v_{\rho}} \right)
              \nonumber \\
& \hspace{1mm}
         + \cosh \left( \frac{(L-\xi) \omega}{v_{\rho}} \right)
         - \cosh \left( \frac{\xi \omega}{v_{\rho}} \right) -1 \Bigg\} , \\
   \label{eq:bar-J-sigma-A}
 \bar{J}_{\sigma}(\omega)
& = \frac{1}{2 \omega \sinh \left( \frac{L \omega}{v_{\sigma}} \right)}
    \Bigg\{  \cosh \left( \frac{L \omega}{v_{\sigma}} \right)
              \nonumber \\
& \hspace{1mm}
         - \cosh \left( \frac{(L-\xi) \omega}{v_{\sigma}} \right)
         - \cosh \left( \frac{\xi \omega}{v_{\sigma}} \right) +1 \Bigg\} , \\
   \label{eq:tilde-J-sigma-A}
 \tilde{J}_{\sigma}(\omega)
& = \frac{1}{2 \omega \sinh \left( \frac{L \omega}{v_{\sigma}} \right)}
    \Bigg\{  \cosh \left( \frac{L \omega}{v_{\sigma}} \right)
              \nonumber \\
 & \hspace{1mm}
         - \cosh \left( \frac{(L-\xi) \omega}{v_{\sigma}} \right)
         + \cosh \left( \frac{\xi \omega}{v_{\sigma}} \right) -1 \Bigg\} .
\end{align}
Integrating out 
$\bar{\lambda_{\rho}}$, $\tilde{\lambda_{\rho}}$, $\bar{\lambda_{\sigma}}$
and $\tilde{\lambda_{\sigma}}$,
we find that
\begin{align}
    \label{eq:intro-S-eff-NZ}
 &  \int \mathcal{D}\bar{\lambda_{\rho}}\mathcal{D}\tilde{\lambda_{\rho}}
         \mathcal{D}\bar{\lambda_{\sigma}}\mathcal{D}\tilde{\lambda_{\sigma}}
    \int \prod_{q>0}
           \mathcal{D}a_{q}^{\dagger} \mathcal{D}a_{q}
           \mathcal{D}b_{q}^{\dagger} \mathcal{D}b_{q}
         \nonumber \\
 & \hspace{37mm} \times
           \exp \left( - S^{\rm NZ} + {\rm i} \mathcal{P} \right)
         \nonumber \\
 & \hspace{5mm}
   \propto \exp \left( - S_{\rm eff}^{\rm NZ} \right) ,
\end{align}
where
\begin{align}
   \label{eq:S-eff-NZ-A}
  S_{\rm eff}^{\rm NZ}
 & = \frac{1}{2 \pi K_{\rho} \beta} \sum_{\omega}
     \bar{J}_{\rho}^{-1}(\omega)
     \bar{\theta}(\omega) \bar{\theta}(-\omega)
         \nonumber \\
 & \hspace{5mm}
   + \frac{1}{8 \pi K_{\rho} \beta} \sum_{\omega}
     \tilde{J}_{\rho}^{-1}(\omega)
     \tilde{\theta}(\omega) \tilde{\theta}(-\omega)
         \nonumber \\
 & \hspace{5mm}
   + \frac{1}{2 \pi K_{\sigma} \beta} \sum_{\omega}
     \bar{J}_{\sigma}^{-1}(\omega)
     \bar{\phi}(\omega) \bar{\phi}(-\omega)
         \nonumber \\
 & \hspace{5mm}
   + \frac{1}{8 \pi K_{\sigma} \beta} \sum_{\omega}
     \tilde{J}_{\sigma}^{-1}(\omega)
     \tilde{\phi}(\omega) \tilde{\phi}(-\omega) .
\end{align}
We thus find that
\begin{align}
   \label{eq:pf-NZ4}
   Z^{\rm NZ}
 & \propto 
    \int \mathcal{D}\bar{\theta}\mathcal{D}\tilde{\theta}
         \mathcal{D}\bar{\phi}\mathcal{D}\tilde{\phi}
         \nonumber \\
 & \hspace{5mm} \times
   \exp \left( - S_{\rm eff}^{\rm NZ}\bigl[ \bar{\theta}, \tilde{\theta},
                                    \bar{\phi}, \tilde{\phi} \bigr]
             - S^{\rm B}\bigl[ \bar{\theta}, \tilde{\theta},
                                    \bar{\phi}, \tilde{\phi} \bigr]
      \right) .
\end{align}
It is clear that the effective action given in eq.~(\ref{eq:S-eff-NZ-A})
satisfies eq.~(\ref{eq:def-S-eff-NZ}).

\section{Derivation of $S^{\Gamma}$}

We assume that our system is symmetric with respect to the left contact
at $x = 0$ and the right contact at $x = L$.
For simplicity, we treat only the coupling at the left contact
between the TL liqid and the left superconductor,
so that we do not explicitly express the subscript $\rm L$,
indicating the left superconductor, in the following derivation.

We first rewrite the tunneling Hamiltonian given in eq.~(\ref{eq:h-t-d})
using the Bogoliubov transformation,
\begin{align}
  H^{\rm T}
 & = \frac{1}{\sqrt{V}} \sum_{k, s}
         \nonumber \\
 & \hspace{1mm} \times
     \Biggl[ d_{k,+}^{\dagger}
              \Big(  u_{k} \int {\rm d}x t_{k}(x) \psi_{+}(x)
         \nonumber \\
 & \hspace{15mm}
                    + v_{k} {\rm e}^{{\rm i}\chi_{1}}
                            \int {\rm d}x t_{-k}^{*}(x)
                                                    \psi_{-}^{\dagger}(x)
             \Big)
         \nonumber \\
 & \hspace{5mm}
          + d_{-k,-}^{\dagger}
              \Big(  u_{k} \int {\rm d}x t_{-k}(x) \psi_{-}(x)
         \nonumber \\
 & \hspace{15mm}
                    - v_{k} {\rm e}^{{\rm i}\chi_{1}}
                            \int {\rm d}x t_{k}^{*}(x)
                                                    \psi_{+}^{\dagger}(x)
             \Big)
         \nonumber \\
 & \hspace{5mm}
          +  \Big(  v_{k} {\rm e}^{-{\rm i}\chi_{1}}
                           \int {\rm d}x t_{-k}(x) \psi_{-}(x)
         \nonumber \\
 & \hspace{15mm}
                   + u_{k} \int {\rm d}x t_{k}^{*}(x)
                                                    \psi_{+}^{\dagger}(x)
             \Big) d_{k,+}
         \nonumber \\
 & \hspace{5mm}
          +  \Big(- v_{k} {\rm e}^{-{\rm i}\chi_{1}}
                           \int {\rm d}x t_{k}(x) \psi_{+}(x)
         \nonumber \\
 & \hspace{15mm}
                   + u_{k} \int {\rm d}x t_{-k}^{*}(x)
                                                    \psi_{-}^{\dagger}(x)
             \Big) d_{-k,-} \Biggr] .
\end{align}
The action $S^{\Gamma}$ is obtained by integrating out the electron field
in the left superconductor,
\begin{equation}
  \exp \left( - S^{\Gamma} \right) \propto
       \int \prod_{k, s} \mathcal{D}d_{k, s}^{\dagger} \mathcal{D}d_{k, s}
           \exp \left( - S^{\rm S} - S^{\rm T} \right) ,
\end{equation}
where
\begin{align}
 S^{\rm S} & = \frac{1}{\beta} \sum_{\epsilon}
                     \left( - {\rm i}\epsilon + E_{k} \right)
      \Big(  d_{k, +}^{\dagger}(\epsilon) d_{k, +}(\epsilon)
         \nonumber \\
 & \hspace{30mm}
            + d_{-k, -}^{\dagger}(\epsilon) d_{-k, -}(\epsilon) \Big) , \\
 S^{\rm T} & = \int_{0}^{\beta} {\rm d}\tau H^{\rm T}(\tau) .
\end{align}
Here, $\epsilon$ denotes the fermion Matsubara frequency.
After the integration, we find that
\begin{align}
  \label{eq:s-gamma-00}
 S^{\Gamma}
 & = \frac{1}{\beta} \sum_{\epsilon} \frac{1}{V} \sum_{k}
                          \frac{u_{k} v_{k}}{- {\rm i}\epsilon + E_{k}}
          \nonumber \\
 & \hspace{1mm} \times
         \biggl(  {\rm e}^{-{\rm i}\chi_{1}}
                  \int {\rm d}x t_{-k}(x) \psi_{-}(x, - \epsilon)
                  \int {\rm d}y t_{k}(y) \psi_{+}(y, \epsilon)
                   \nonumber \\
 & \hspace{5mm}
               + {\rm e}^{{\rm i}\chi_{1}}
                  \int {\rm d}x t_{k}^{*}(x)
                                       \psi_{+}^{\dagger}(x, \epsilon)
                  \int {\rm d}y t_{-k}^{*}(y)
                                       \psi_{-}^{\dagger}(y, - \epsilon)
                   \nonumber \\
 & \hspace{5mm}
               - {\rm e}^{-{\rm i}\chi_{1}}
                  \int {\rm d}x t_{k}(x) \psi_{+}(x, - \epsilon)
                  \int {\rm d}y t_{-k}(y) \psi_{-}(y, \epsilon)
                   \nonumber \\
 & \hspace{5mm}
               - {\rm e}^{{\rm i}\chi_{1}}
                  \int {\rm d}x t_{-k}^{*}(x)
                                       \psi_{-}^{\dagger}(x, \epsilon)
                  \int {\rm d}y t_{k}^{*}(y)
                                       \psi_{+}^{\dagger}(y, - \epsilon)
         \biggr) .
\end{align}
In deriving eq.~(\ref{eq:s-gamma-00}), we neglected irrelevant terms
which do not depend on the macrosocpic phase $\chi_{1}$.
Equation (\ref{eq:s-gamma-00}) is simplified to
\begin{align}
    \label{eq:s-gamma-11}
 S^{\Gamma}
 & = \frac{1}{\beta} \sum_{\epsilon}
                  \frac{1}{V} \sum_{k}
                      \frac{2 E_{k}}{\epsilon^{2} + E_{k}^{2}} u_{k}v_{k}
         \nonumber \\
 & \hspace{5mm} \times
         \biggl(  {\rm e}^{-{\rm i}\chi_{1}}
                  \int {\rm d}x \int {\rm d}y t_{-k}(x) t_{k}(y)
                      \psi_{-}(x, - \epsilon) \psi_{+}(y, \epsilon)
         \nonumber \\
 & \hspace{50mm}
                     + {\rm h. c.} \biggr) .
\end{align}

We assume that the tunneling-matrix element satisfies
\begin{equation}
    \label{eq:t-t-ave}
  \langle t_{k}(x) t_{-k}(y) \rangle_{k_{\rm F}}
   = \tilde{\Gamma} (x) \delta (x - y) ,
\end{equation}
where $\langle \cdots \rangle_{k_{\rm F}}$ denotes the average
over $k$ on the Fermi surface.
Note that $\tilde{\Gamma}(x)$ has nonzero values only in the vicinity
of $x = 0$ and vanishes when $\alpha_{\rm T} < x$.
If we carry out the integrations over $x$ and $y$ in eq.~(\ref{eq:s-gamma-11})
after averaging over $k$ on the Fermi surface,
the cross terms between $\psi_{1\pm}(x)$ and $\psi_{2\mp}(x)$ vanish
due to the presence of a spatially oscillating factor of
${\rm e}^{\pm {\rm i} 2 k_{\rm F}x}$.
After the summation over $k$, we find that
\begin{align}
 S^{\Gamma}
 & \approx \frac{1}{\beta} \sum_{\epsilon}
              \frac{\Gamma \Delta}
                      {\sqrt{\Delta^{2} + \epsilon^{2}}}
          \biggl( {\rm e}^{-{\rm i}\chi_{1}}
                   \Big(  \psi_{1-}(0, - \epsilon) \psi_{2+}(0, \epsilon)
         \nonumber \\
 & \hspace{20mm}
                         + \psi_{2-}(0, - \epsilon) \psi_{1+}(0, \epsilon)
                   \Big)
                     + {\rm h. c.} \biggr) ,
\end{align}
where $\Gamma = \pi N(0) \int {\rm d}x \tilde{\Gamma}(x)$
($N(0)$: density of states at the Fermi level).
In the imaginary-time representation,
$S^{\Gamma}$ is rewritten as
\begin{align}
  S^{\Gamma}
 & = \int_{0}^{\beta} {\rm d}\tau_{1} {\rm d}\tau_{1}
                Q(\tau_{1} - \tau_{2})
          \biggl( {\rm e}^{-{\rm i}\chi_{1}}
                   \Big(  \psi_{1-}(0, \tau_{1}) \psi_{2+}(0, \tau_{2})
         \nonumber \\
 & \hspace{20mm}
                         + \psi_{2-}(0, \tau_{1}) \psi_{1+}(0, \tau_{2})
                   \Big)
                     + {\rm h. c.} \biggr) ,
\end{align}
where
\begin{equation}
 Q(\tau) = \frac{1}{\beta} \sum_{\epsilon}
             \frac{\Gamma \Delta}{\sqrt{\Delta^{2} + \epsilon^{2}}}
             {\rm e}^{- {\rm i} \epsilon \tau} .
\end{equation}
Note that we are interested in the case of $T \ll \Delta$,
where $Q(\tau)$ is approximated as
\begin{equation}
 Q(\tau) = \frac{1}{\pi} \int_{\Delta}^{\infty} {\rm d}z
             \frac{\Gamma \Delta}{\sqrt{z^{2} - \Delta^{2}}}
             {\rm e}^{- | \tau | z} .
\end{equation}
This indicates that $Q(\tau)$ decays exponentially with a characteristic
time scale of the order of $\Delta^{-1}$.

Adding the term describing the coupling at the right contact,
we finally obtain
\begin{align}
  S^{\Gamma}
 & = \int_{0}^{\beta} {\rm d}\tau_{1} {\rm d}\tau_{1}
                Q(\tau_{1} - \tau_{2})
         \nonumber \\
 & \hspace{5mm} \times
          \biggl(  {\rm e}^{-{\rm i}\chi_{1}}
                    \Big(  \psi_{1-}(0, \tau_{1}) \psi_{2+}(0, \tau_{2})
         \nonumber \\
 & \hspace{12mm}
                          + \psi_{2-}(0, \tau_{1}) \psi_{1+}(0, \tau_{2})
                    \Big)
                      + {\rm h. c.}
             \nonumber \\
 & \hspace{12mm}
                 + {\rm e}^{-{\rm i}\chi_{2}}
                    \Big(  \psi_{1-}(L, \tau_{1}) \psi_{2+}(L, \tau_{2})
         \nonumber \\
 & \hspace{12mm}
                          + \psi_{2-}(L, \tau_{1}) \psi_{1+}(L, \tau_{2})
                    \Big)
                      + {\rm h. c.}                
          \biggr) .
\end{align}

\end{document}